\documentclass[conference]{IEEEtran}
\usepackage{cite}
\usepackage{amsmath,amssymb,amsfonts}
\usepackage{algorithmic}
\usepackage{graphicx}
\usepackage[justification=centering]{caption}
\usepackage{xcolor}

\usepackage{booktabs} 
\usepackage{subfig}
\usepackage{amsmath}
\usepackage[normalem]{ulem}
\usepackage{enumitem}
\usepackage{multirow}
\useunder{\uline}{\ul}{}
\usepackage[colorlinks,linkcolor=blue]{hyperref}
\usepackage[nomargin,inline,marginclue,draft]{fixme}
\usepackage{balance}
\usepackage{changepage}

\newtheorem{definition}{Definition}

\newlength\savedwidth

\newcommand{\para}[1]{{\vspace{4pt} \bf \noindent #1 \hspace{10pt}}}

\def\BibTeX{{\rm B\kern-.05em{\sc i\kern-.025em b}\kern-.08em
    T\kern-.1667em\lower.7ex\hbox{E}\kern-.125emX}}
\begin{document}

\newcommand{\authordrift}{\hspace{0.45cm}}

\title{Inhomogeneous Social Recommendation with Hypergraph Convolutional Networks}

\IEEEoverridecommandlockouts
\author{\IEEEauthorblockN{Zirui Zhu}
	\IEEEauthorblockA{\textit{Department of Electronic Engineering} \\
		\textit{Tsinghua University,}\\
		Beijing, China \\
			zhuzr17@mails.tsinghua.edu.cn}
	\and
\IEEEauthorblockN{Chen Gao}
\IEEEauthorblockA{\textit{Department of Electronic Engineering} \\
	\textit{Tsinghua University,}\\
	Beijing, China \\
	chgao96@gmail.com}
\and
\IEEEauthorblockN{Xu Chen$^*$\thanks{$^*$ Corresponding author.}}
\IEEEauthorblockA{
    \textit{Beijing Key Laboratory of Big Data}\\
    \textit{Management and Analysis Methods,}\\
    \textit{Gaoling School of Artificial Intelligence}\\
	\textit{Renmin University of China,}\\
	successcx@gmail.com}
\and
\IEEEauthorblockN{\authordrift Nian Li}
\IEEEauthorblockA{\authordrift \textit{Department of Electronic Engineering} \\
	\authordrift \textit{Tsinghua University,}\\
	\authordrift Beijing, China \\
	\authordrift lin17@mails.tsinghua.edu.cn}
\and
\IEEEauthorblockN{Depeng Jin}
\IEEEauthorblockA{\textit{Department of Electronic Engineering} \\
	\textit{Tsinghua University,}\\
	Beijing, China \\
	jindp@tsinghua.edu.cn}
\and
\IEEEauthorblockN{Yong Li}
\IEEEauthorblockA{\textit{Department of Electronic Engineering} \\
	\textit{Tsinghua University,}\\
	Beijing, China \\
	liyong07@tsinghua.edu.cn}
}

% 	\author{
% 	\IEEEauthorblockN{
% 	Zirui~Zhu\IEEEauthorrefmark{2},
% 	Chen~Gao\IEEEauthorrefmark{2},
% 	Nian~Li\IEEEauthorrefmark{2},
% 	Xu~Chen\IEEEauthorrefmark{3},
% 	Depeng~Jin\IEEEauthorrefmark{2},
% 	Yong~Li\IEEEauthorrefmark{2}
% 	}
% 	\IEEEauthorblockA{
% 	\IEEEauthorrefmark{2}Department of Electronic Engineering, Tsinghua University, Beijing, China\\
% 	\IEEEauthorrefmark{3}Gaoling School of Artificial Intelligence, Renmin University of China, Beijing, China\\
%     }
%     }

% \markboth{IEEE TRANSACTIONS ON KNOWLEDGE AND DATA ENGINEERING}
% {Shell \MakeLowercase{\textit{et al.}}: Bare Demo of IEEEtran.cls for Computer Society Journals}
% %{Shell \MakeLowercase{\textit{et al.}}: Bare Demo of IEEEtran.cls for Computer Society Journals}

\begin{comment}
1. application, uui
(First, the impact of this paper may be limited. As the author mentions, public dataset usually only has a “user-user” relation or “user-item” record, and only a few datasets have a “user-user-item” relation. Therefore, it seems that this hypergraph method cannot be widely used in the real world.)
2. why existing methods fail to model social influence.
(The collaborative filtering approach could capture such latent and implicit relationships. Nevertheless, I noted that authors have attempted to model this explicitly. Perhaps the authors may want to motivate this better to showcase that existing graph-based deep recommender systems are not able to capture such implicit relationships.)
3. typos and grammar mistakes
\end{comment}

\IEEEtitleabstractindextext{

\begin{abstract}
Incorporating social relations into the recommendation system, \textit{i.e.} social recommendation, has been widely studied in academic and industrial communities.
While many promising results have been achieved, existing methods mostly assume that the social relations can be homogeneously applied to \textit{all the items}, which is not practical for users' actually diverse preferences.
In this paper, we argue that the effect of the social relations should be inhomogeneous, that is, two socially-related users may only share the same preference on some specific items, while for the other products, their preferences can be inconsistent or even contradictory.
Inspired by this idea, we build a novel social recommendation model, where the traditional pair-wise ``user-user'' relation is extended to the triple relation of ``user-item-user''.
To well handle such high-order relationships, we base our framework on the hypergraph.
More specifically, each hyperedge connects a user-user-item triplet, representing that the two users share similar preferences on the item. We develop a \textbf{S}ocial \textbf{H}yper\textbf{G}raph \textbf{C}onvolutional \textbf{N}etwork (short for \textbf{SHGCN}) to learn from the complex triplet social relations.
With the hypergraph convolutional networks, the social relations can be modeled in a more fine-grained manner, which more accurately depicts real users' preferences, and benefits the recommendation performance.
Extensive experiments on two real-world datasets demonstrate our model's effectiveness. Studies on data sparsity and hyper-parameter studies further validate our model's rationality. Our codes and dataset is available at https://github.com/ziruizhu/SHGCN.
\end{abstract}


}

\maketitle
\IEEEdisplaynontitleabstractindextext
\IEEEpeerreviewmaketitle

\section{Introduction}
As an effective remedy for information overloading, the recommender system has been deployed in a multitude of real-world applications.
With the rapid development of online social networks, how to better exploit social relations for recommender system has gained its increasing popularity and various methods have emerged.
The major approaches can be divided into two categories.
Some works~\cite{soreg,jamali2010matrix,wang2017item} propose to use regularization methods or multi-task learning to make the distance between friends as short as possible.
Some other works~\cite{sorec,yang2017bridging,wu2019neural} propose to smooth the friends' embedding by sharing latent representations between friends.

%%%%%%% <<< NEW VERSION
Nevertheless, these existing works have ignored the significant fact that users share inhomogeneous interests with friends, resulting in inferior recommendation performance. For example, a user may share similar interests in books with classmates and share the same taste with family members on food or dressing. 
In other words, social relations have an inhomogeneous influence on users' behaviors. 
It is worth mentioning that there are some works of social recommendation modeling the various strength of social relations~\cite{DiffNet++,SAMN,DualGAT}. However, such strength only represents the extent of social closeness but cannot handle the inhomogeneous influence\footnote{These works also use the term ``influence", but however they roughly use strength to represent the influence.}. 

\begin{figure}[t]
    \begin{center}
    \mbox{
        \subfloat[Demo of product sharing]{\label{Fig:hyperedgedemo:a}\includegraphics[width= 4cm]{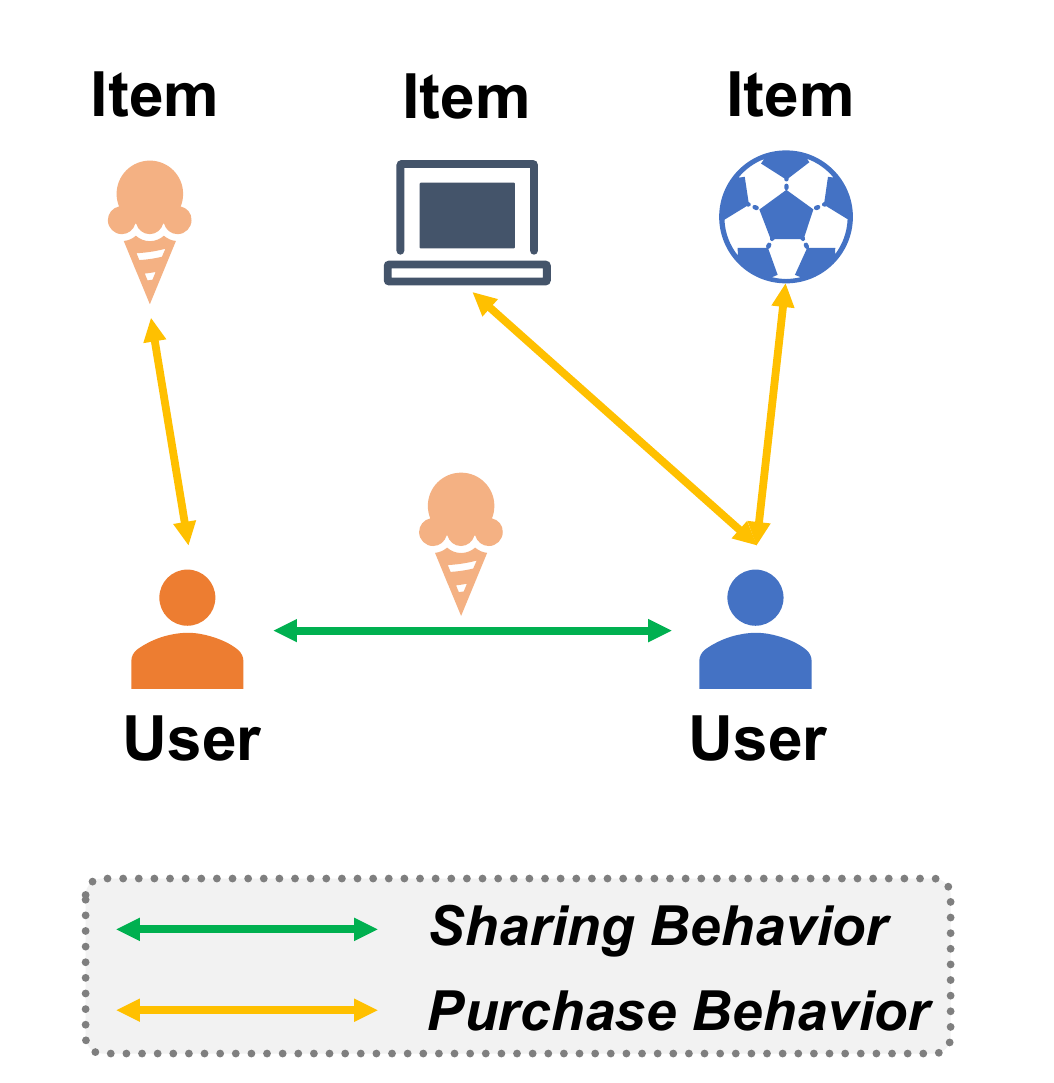}}
        \hspace{0.3cm}
        \subfloat[Demo of group buying]{\label{Fig:hyperedgedemo:b}\includegraphics[width=4cm]{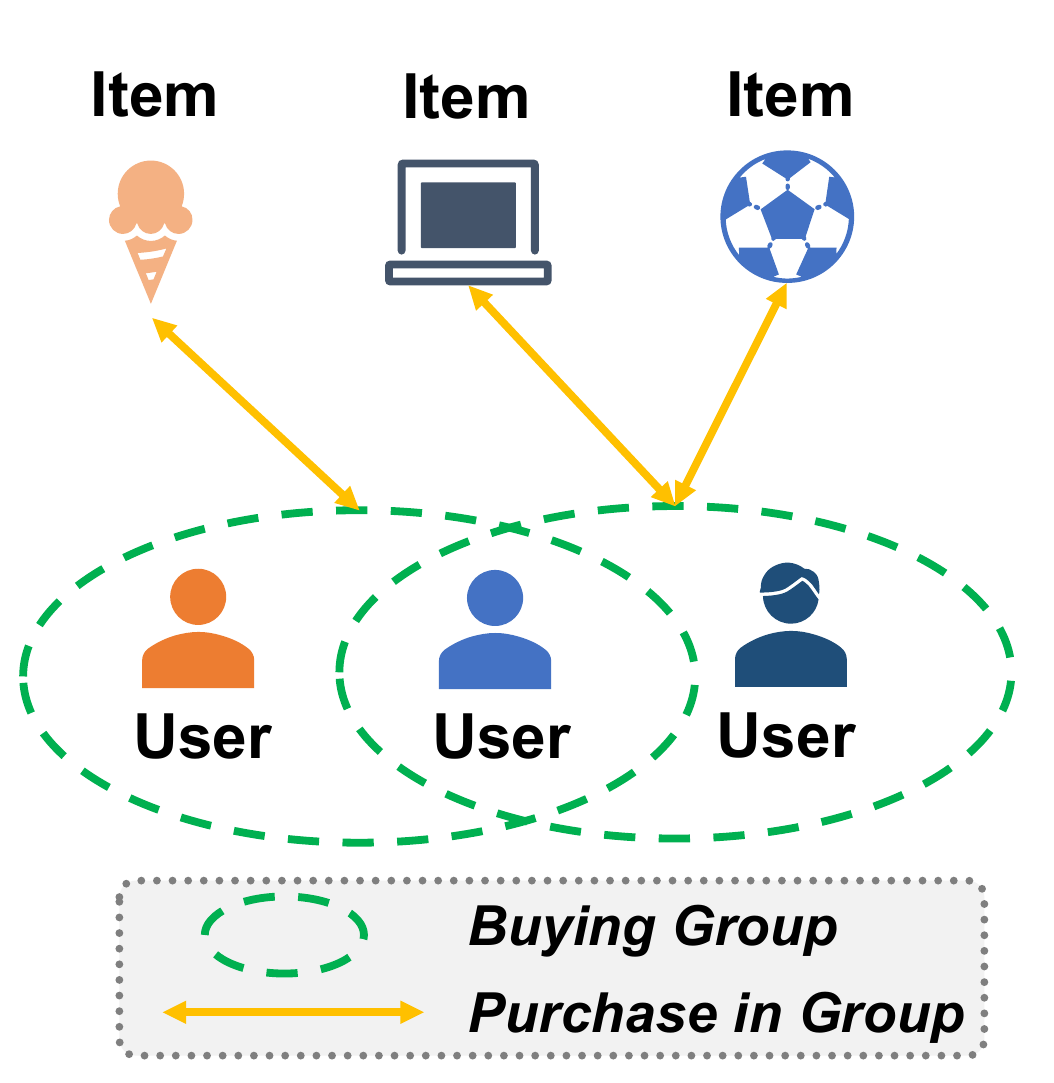}}
    }
    \end{center}
    \caption{Two examples of triplet social relation}\label{Fig:hyperedgedemo}
\end{figure}

Such inhomogeneous effects are of importance but hard to capture, since at most times we can only obtain the two-tuple social relations.
Recently, social e-commerce platforms such as Pinduoduo.com are gaining popularity. In this new kind of e-commerce platform, users can share products with their friends on the social network, as is illustrated in Figure \ref{Fig:hyperedgedemo:a}. When a user shares an item with his/her friend, the shared product can reflect the fine-grained common interests between them, to some extent. 
Another case in this platform is group-buying where two (or more) users launch a buying group and buy a specific item together, as is shown in Figure \ref{Fig:hyperedgedemo:b}. The inhomogeneous effect plays an important role here, considering that people would buy different items with different friends, \textit{e.g.} rackets with tennis buddies and laptops with colleagues. 

These behaviors provide us a precious opportunity for studying the inhomogeneous social influence on user behaviors. It is not difficult to summarize the preceding examples as triple relation of ''user-user-item'' in a unified form. 
However, modeling the triple relation of ``user-user-item'' is seldom explored by existing works. Directly considering inhomogeneous social relations as homogeneous or giving them unidimensional weights cannot represent the inhomogeneous social relations instinctively. There are two main challenges,
\begin{itemize}[leftmargin=*]
    \item \textbf{Representation of the triple social relations.} The triple social relations involve three sides, two users and a shared item. The relations between them are not clear, and it is quite challenging to construct the representation, compared with existing works where there is only a scalar value for representing the pairwise relation of two users.
    \item \textbf{Exploiting the triplets in preference learning.} The triplets reflect complex and fine-grained common interests between two users, indeed. Given the triplets, it is challenging to distill the prediction signal and fuse it into preference learning. 
\end{itemize}
Inspired by the recent advances in graph learning ~\cite{feng2019hypergraph,ijcai2019-366}, we propose to construct a hypergraph, which generalizes the graph by introducing hyperedges that can connect more than two nodes.
More precisely, we utilize hyperedges to connect two user nodes and an item node for representing complex triplets. To capture complex social influence and learn user preferences, we propose a hypergraph convolutional network-based model, named SHGCN. With the carefully designed embedding-propagation layers, the model can effectively learn users' latent preferences through messages passing on the hypergraph constructed by the triple social relations.

Our contribution can be summarized as follows,
\begin{itemize}[leftmargin=*]
    \item We approach the problem of social recommendation from a novel perspective of inhomogeneous social influence. In this setting, social influence is modeled from a fine-grained perspective, which is more general compared with traditional social recommendation.
    \item We propose to construct a hypergraph that can represent both the complex triple social-relations and user-item interaction data. 
    Specifically, the hyperedges on the hypergraph can well encode the triple social-relations.
    We then propose a hypergraph convolutional network-based model to capture the inhomogeneous social influence and user preference for recommendation.
    \item We conduct experiments on two real-world datasets to evaluate our proposed model. The empirical results demonstrate that our model can outperform the state-of-the-art baselines by 2.18\% to 13.26\%. Further studies confirm our model's effectiveness for both sparse users. We also find that our model is not sensitive to various hyper-parameter settings, verifying its high application value in the real world.
\end{itemize}

The remainder of this paper is as follows. We first formulate our problem in Section~\ref{sec::probdef} and present our solution in Section~\ref{sec::method}. We then conduct experiments in Section~\ref{sec::exp} and review the related works in Section~\ref{sec::related}. Last, we conclude our paper and discuss future works in Section~\ref{sec::conclusion}.

\section{Problem Formulation}\label{sec::probdef}
\begin{figure*}[!t]
	\begin{center}
		\mbox{
			\subfloat{\includegraphics[scale=0.5]{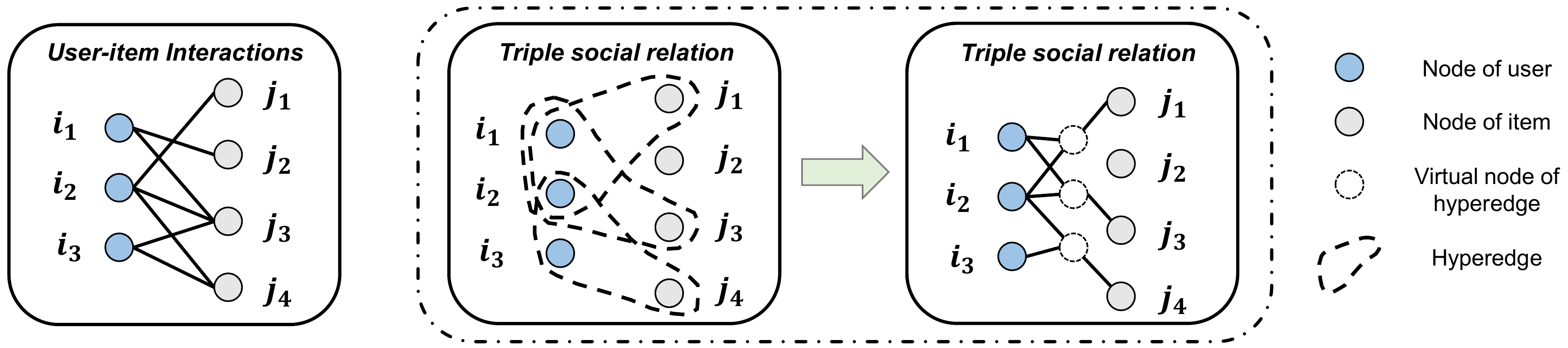}}
		}
	\end{center}
	\caption{Illustration of how we use hyperedge to represent the complex triple social relations.}\label{fig::demoHyper}
\end{figure*}

The traditional social recommendation is defined as to recommend based on user-item interaction data and binary social-relation paired data.
Different from it, in inhomogeneous social recommendation, the social-relation data is in the triple form, which is illustrated in Figure \ref{Fig:hyperedgedemo}.
For example, a user can share an item to or co-purchase an item with his/her friends, which reflects their fine-grained common interests. 
Generally, the social-relation can be represented as $<$user, user, item$>$.

Then the problem of inhomogeneous social recommendation turns to recommend with user-item interaction data and triple social-relation paired data.
Assume that the set of users/items is $\mathcal{U}$/$\mathcal{V}$ and there are $M$ users and $N$ items. The user-item interaction data can be denoted as a set $\mathcal{Y}$, defined as follows,
\begin{equation}
    \mathcal{Y}= \{(i,j)|\text{user $i$ interacts with the item $j$}  \}.
\end{equation}
The triple social-relation data can be denoted as a set of triplet $\mathcal{E}$, defined as follows,
\begin{equation}
\small
   \mathcal{E} = \left\lbrace e=(i_1,i_2,j), i_1,i_2\in \mathcal{U}, j \in \mathcal{V} \;\middle|\;
  \begin{tabular}{@{}l@{}}
    user $i_1$ interacts with friend $i_2$ \\with respect to item $j$
   \end{tabular}
  \right\rbrace
\end{equation}

Then the studied problem in this work can be formulated as follows,

\textbf{Input}: User-item interaction data $\mathcal{Y}$ and triple social-relation data $\mathcal{E}$.

\textbf{Output}: A recommendation model that can estimate the probability that a user $i$ will interact (purchase) with an item $j$.

With the obtained model, we can rank all item candidates according to the predicted scores and then select the top-ranked items as the recommendation results.
\section{Methodology}\label{sec::method}
% We propose \textbf{S}ocial \textbf{H}yper\textbf{G}raph \textbf{C}onvolutional \textbf{N}etwork (short for \textbf{SHGCN}) and 
Our proposed \textbf{S}ocial \textbf{H}yper\textbf{G}raph \textbf{C}onvolutional \textbf{N}etwork (short for \textbf{SHGCN}) can be summarized as four parts as follows,
\begin{itemize}[leftmargin=*]
\item \textbf{Hypergraph Construction:}
To better model the triplet social relation that involves multiple users and items, we first construct a hypergraph,
which generalizes the concept of edge in existing graph-based models.
The traditional graph-based models then can be regarded as a de-generated case where all edges’ degrees are two, also called two-uniform hypergraph. 
Then triple social relation can be modeled as hyperedge with a degree equal to three.

 \item \textbf{Embedding Layer:}
For each node of the constructed hypergraph, we assign it with a trainable low-dimensional embedding vector in the latent space, which is the fundamental of the following hypergraph convolutional layer. Here we consider the users or items as the same to ensure they are represented in the same space.
%  Following the mainstream recommendation systems, we represent a node with a trainable low-dimensional embedding vector regardless of whether it is user or item.

 \item \textbf{Hypergraph Convolutional Layer:}
 To capture the inhomogeneous social influence through the triplet social relation represented as hyperedge, we propose the hypergraph convolutional layers that propagate embeddings on the hypergraph, making the embeddings of both vertexes and hyperedges can absorb in the neighbours' information.
% Hyperedge modeling and skip connections established between users for more effective message passing are the core conceptions of this part. It is worth noting that specific propagation path might be designed between item and user, item and item, according to the inherent characteristics of different datasets.

 \item \textbf{Prediction and Optimization:}
%  Our primary focus is to establish embedding learning algorithms, so we take the most commonly used rating function,
 To obtain recommendation results, we utilize the simple yet effective \textit{i.e.} inner product, which is also very efficient demonstrated by existing works~\cite{rendle2020kdd,wang2019neural}.
%  , despite the existence of other potential choices. 
We deploy the widely-used Bayesian Personalized Ranking (BPR) loss \cite{rendle2009bpr} to optimize the model parameters.
\end{itemize}

\begin{table}[t!]
\centering
\caption{Commonly used notations}
\label{tab:notation}
\begin{tabular}{c|c}
		\hline
\textbf{Symbols} & \textbf{Descriptions} \\ \hline
$\mathcal{U}/\mathcal{V}$ & The set of users/items \\ \hline
$M/N$ & The number of users/items\\ \hline
$\mathcal{X}$ & The set of nodes \\ \hline
$\mathcal{E}$ & The set of hyperedges \\ \hline
$\mathcal{Y}$ & The user-item interaction set \\ \hline
$\mathcal{T}$ & The set of social relations \\ \hline
$i/j$ & User/Item's ID \\ \hline
$e/t$ & Hyperedge's/Relation's ID \\ \hline
% $w$ & General ID of nodes or hyperedges \\ \hline
$r_{ij}$        & The prediction value of item $j$ by user $i$ \\ \hline
$\mathbf{P}_i/\mathbf{Q}_j$        & The trainable embedding of user $i$ / item $j$  \\ \hline
$\mathbf{C}_e$        & The derivative embedding for the hyperedge $e$ \\ \hline
$\mathbf{R}_t$        & The derivative embedding for the relationship $t$ \\ \hline
$\mathbf{E}$          & The embedding of matrix of users and items \\ \hline
$(\cdot)^k$  & The output of the $k$-th layer \\ \hline
$d$        & The length of embedding vector \\ \hline
$\eta(i_1, i_2)$       & \begin{tabular}[c]{@{}c@{}}   The function mapping two connected user $i_1$ and $i_2$ \\to their relation's index \end{tabular}                           \\ \hline

% $\mathcal{B}(i)$       &  The set of items which user $i$ purchased\\ \hline
% $\mathcal{D}(j)$       &  The set of users who purchased item $j$ \\ \hline
$\mathcal{Z}(\cdot)$   &  The set of hyperedges connected to the input node \\ \hline
$\mathcal{K}(e)$       &  The set of nodes connecting to hyperedge $e$ \\ \hline
$\mathcal{N}(i)$       &  \begin{tabular}[c]{@{}c@{}} The set of social friends who are\\ connected to user $i$ by hyperedge(s)  \end{tabular} \\ \hline
$\mathcal{N}(i_1, i_2)$  &  \begin{tabular}[c]{@{}c@{}} The set of hyperedges connecting to\\both user $i_1$ and user $i_2$  \end{tabular} \\ \hline

% $L_c$       & \begin{tabular}[c]{@{}c@{}} The adjacent matrix between\\hyperedges and nodes \end{tabular}\\ \hline
% $D_c$       &  The diagonal degree matrix of $L_c$\\ \hline
% $L_r$& \begin{tabular}[c]{@{}c@{}} The adjacent matrix between\\relationshios and hyperedges \end{tabular} \\ \hline
% $D_r$       &  The diagonal degree matrix of $L_r$\\ \hline

% $L_i$ & \begin{tabular}[c]{@{}c@{}} The adjacent matrix of \\ the user-item bipartite \end{tabular} \\ \hline
% $D_i$       &  The diagonal degree matrix of $L_i$\\ \hline

$\alpha _{i_1i_2}^k$         &   \begin{tabular}[c]{@{}c@{}}The user attention of user $i_1$ in \\contributing to $\mathbf{p}_{i_2}^k$   \end{tabular}                              \\ \hline
$|\cdot|$ & Cardinality of the set or the norm of vector. \\ \hline
$\cdot||\cdot$ & Concatenation of two vectors \\ \hline
$\sigma(\cdot)$ & The activation function \\ \hline
$\otimes$     &  The inner prodct function \\ \hline
$\text{MLP}(\cdot)$ & Multilayer perceptron \\\hline
$W, b$ & The weight and bias in neural network \\\hline
% \hline
	\end{tabular}
\end{table}
 
\subsection{Hypergraph Construction}
\subsubsection{Hyperedge and Hypergraph}
Firstly, we give a brief introduction of \textit{hypergraph}. Hypergraph generalizes the classical graph that only models pairwise relations between objects by replacing edge with \textit{hyperedge}. A hyperedge can connect any number of vertices. The formal definition can be summarized as follows,

\begin{definition}
\textbf{Hypergraph.} A hypergraph $H$ can be defined as $H=(\mathcal{X},\mathcal{E})$, where $\mathcal{X}$ is a set of nodes (also called vertices), and $\mathcal{E}$ is a set of  hyperedge. Each hyperedge connects several vertices, and thus it can be regarded as a non-empty set of vertices.
\end{definition}

A hyperedge can connect any vertices, and thus
a hyperedge $e \in \mathcal{E}$ is an element of $\mathcal{P}\backslash\{\emptyset\}$, where $\mathcal{P}$ denotes the \textit{power set} of $\mathcal{X}$. 
For $\forall e \in \mathcal{E}$, the cardinality of $e$  is also called the degree of $e$. 
Therefore, according to the definition above, hyperedges connecting just two vertices can be regarded as the classical graph edge. 
For the sake of simplicity and precision, in this paper, we refer to hyperedge as the hyperedge with a degree larger than two.

\subsubsection{Hypergraph-structured data}
Traditional social recommendation algorithms cannot well handle triple social relations since they are designed to tackling classical graph-structured data. 
Essentially, the input data of inhomogeneous social recommendation is a hypergraph constructed by all kinds of interactions between users and items as mentioned in section~\ref{sec::probdef}. 
If we must adapt these methods to the hypergraph, 
then we should degrade hyperedges to classical edges, which will make the high-ordered interaction information revealed in the hyperedges discarded.

We represent user and item as nodes, which is a commonly-accepted manner in existing works~\cite{wang2019neural,ying2018graph}.
To construct our data as hyperedge, we first build a hyperedge between user $i_1$, user $i_2$ and item $j$ for any triple social relation $(i_1,i_2,j)\in\mathcal{E}$; for the  user-item interaction $(i,j)\in\mathcal{Y}$, we build a classical edge between user $i$ and item $j$. 
Then the task of inhomogeneous social recommendation turns to predict the existence of classical interaction edge between a given user and a given item on the built hypergraph.

It is worth mentioning that the hyperedges actually can connect any number of users and items. In our problem, we only consider the hyperedge connecting two users and one item, which represents triplet social relations. But no matter how, the proposed manner of hypergraph construction can easily handle more complex relations with more users or items.
% Hyperedges with arbitrary degrees could be integrated into the hypergraph though it is only constructed by 2-degree ordinary edges and 2-degree hyperedges in the scenario of our problem. 
% It is not difficult to imagine that such complex interaction would be more and more common with the rapid development of social e-commerce.

% The input data of recommendation system is mainly the interactions between users and items. Traditional recommendation systems formalize the data as bipartite considering that the most important interactions happen only between user and item. With the rapid development of social e-commerce, the edge between different users has been integrated into the bipartite to capture the social influence on users’ preference. However, existing frameworks still fail to model triple social relation, such as sharing an item to one's friend or participating in a group buying, as these mentioned behaviors are all involved with more than two entities, which does not satisfy the definition of the traditional edge.
% We use hyperedge with degree equal to three, which is connected with two users and one item, to represent item sharing and group buying interactions and regular edges connecting one user and one item to represent the purchase interaction. So in the scenario of our problem, the hypergraph is composed of simple edge with degree equal to two and hyperedge with degree equal to three.
% \chen{clear definition of our hypergraph-structured data. RE:done}

\begin{figure*}[!t]
\vspace{0.4cm}
	\begin{center}
		\mbox{
			% original: width=8.5
		\subfloat{\includegraphics[scale=0.4]{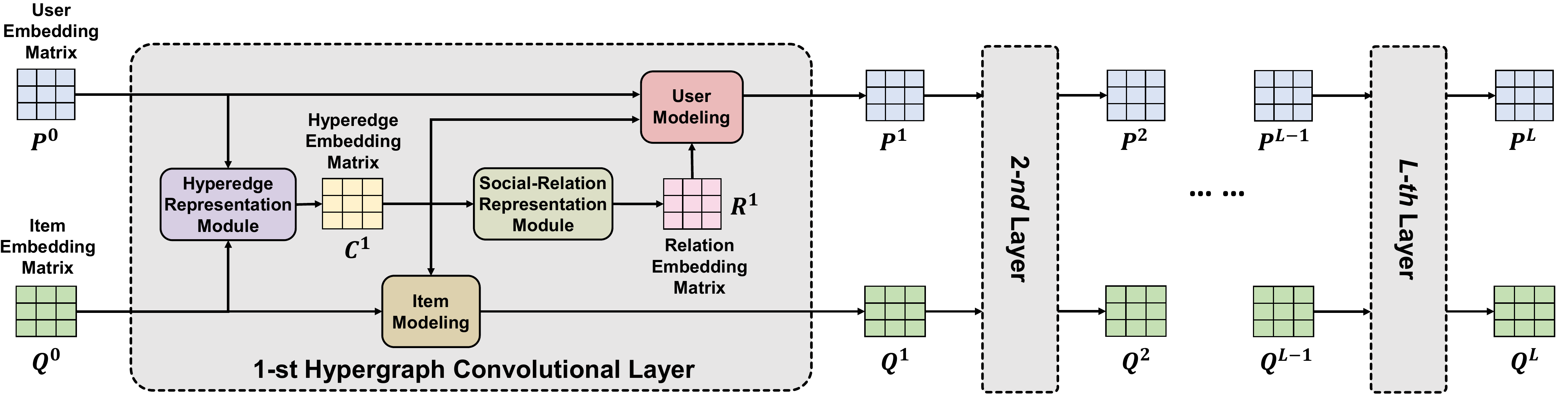}}
		}
	\end{center}
	\caption{Illustration of SHGCN's propagation process in section~\ref{sec::method::hyper}.}\label{fig::prop}
\end{figure*}

\subsection{Embedding Layer}
We describe a user $i$ (an item $j$) with a low-dimensional vector $\mathbf{P}_i{\in}\mathbb{R}^d$ ($\mathbf{Q}_j{\in}\mathbb{R}^d$), where $d$ denotes the embedding size, following the paradigm of existing recommendation models ~\cite{he2017neural,ying2018graph}. Then the full embedding matrix, containing both user and item, can be formulated as follows,
\begin{equation}\label{equation:embedingMatrix}
\small
	\begin{aligned}
    \mathbf{E} = [\mathbf{P}_{1}, \cdots, \mathbf{P}_{{M}}, \mathbf{Q}_{1}, \cdots, \mathbf{Q}_{{N}}] = [\mathbf{P}, \mathbf{Q}],
	\end{aligned}
\end{equation}
where $M$ and $N$ denote the number of users and items, respectively. In the following sections, we use $\mathbf{E}^k{=}[\mathbf{P}^k, \mathbf{Q}^k]$ to represent the embedding matrix obtained by $k$-th hypergraph convolutional layer. We have $\mathbf{E}^0{=}\mathbf{E}$ here.

% \chen{existing models cannot handle... different from them, we xxxx}
% \chen{\textcolor{blue}{Each paragraph has a specific function; no more than 3 lines}}
Existing graph-based methods always only assign trainable embedding matrices to nodes and ignore the explicit representation of edges. In the input data, the hyperedge encodes the triplet social relation and reveals inhomogeneous social influence. Therefore, it is essential to assign representations to these hyperedges. 

% Most of graph-based methods use a trainable embedding matrix to capture the latent factor of nodes and an adjacent matrix or laplacian matrix to represent all edges. Hence, there is a underlying inequality in entropy of nodes and edges.
% We argue that hyperedges deserve more computational resources as removing the restriction of edge's degree brings much more possibilities in the structure of hypergraph. 

However, providing each hyperedge a freely-trainable vector will cost extremely high memory, since the space of hyperedge is huge.
% the effort to give every hyperedge a trainable vector is also doomed in vain, with an out-of-memory error, considering that interactions are generated dynamically. 
Therefore, we seek to make a trade-off that we generate a representative vector $\mathbf{C}_e$ for hyperedges $e$ from the vertices connected by $e$. 
In the following section, we address it via hypergraph convolution operations.
% This part would be fully illustrated in the next section.

% \chen{why we need to propagate? 1) extract triplet's signal in latent space, propagate helps to assign embedding 2) exploit triplet's latent representation in preference learning. one paragraph.}

% \chen{1. hyper-edge representation 2.hyper-edge exploitation 2.1 Social-aware User Modeling para social-relationship representation para user representation 2.2 Item modeling}

\subsection{Hypergraph Convolutional Layer}\label{sec::method::hyper}
In this section, we would give an elaborate description of the whole hypergraph convolutional layer. Firstly, we derive representation of hyperedges to explicitly capture the inhomogeneous social influence that hyperedges reveal. 
The representation of hyperedges will be further exploited to model user-user social relations, users, and items. The overall structure is illustrated in Figure~\ref{fig::prop}. 

\subsubsection{Hyperedge Representation}
% The key idea to tackle the hyperedges is 
To make it easier to understand, we can regard hyperedges as virtual nodes, as is shown in Figure ~\ref{fig::demoHyper}.
For a virtual node of hyperedge, it is adjacent to the nodes it connects as a classical edge. 
% Then the graph with hyperedges is compatible with the ordinary graph in some way.
% Interactions represented by hyperedge are much more complex compared to direct user-item interactions. Take product sharing behaviour as an example. When a user shares the same product with different users, or when two users share different products, the meaning of the sharing behavior itself is different. If we can't capture the node characteristics related to the sharing behavior precisely, we may not be able to correctly evaluate how such sharing behavior affects the information disseminated on the whole network.

Then message propagation of hypergraph convolutional network to obtain the embedding $C_e$ for hyperedge $e$ can be re-formulated as the propagation in graph convolutional network as follows,
\begin{equation}\label{equation:hyperedgeModeling}
\small
	\begin{aligned}
    & \mathbf{C}_e^k = \sigma \left({\textit{aggregate}} \left(\mathbf{E}_w^{k-1} | w\in  \mathcal{K}(e)\right)\right), \\
	\end{aligned}
\end{equation}
where $\mathcal{K}(e)$ denotes the set of nodes connected by hyperedge $e$, $\sigma(\cdot)$ is the activation function, $\mathbf{E}_w^{k-1}$ denotes the embedding matrix at the $(k-1)$-th layer. Note that we do not distinguish the user and item embeddings, and $w$ denotes the uniform index of user or item.

For the $l$-uniform hypergraph, we can concatenate all the embedding of adjacent nodes and then use a multi-layer perceptron (MLP) as an aggregator to generate the embedding of hyperedge.
However, we have observed severe over-fitting issue in experiments. Therefore we choose the simple yet effective way following traditional graph convolutional networks~\cite{wu2019neural} as follows,
\begin{equation}\label{equation:hyperedgeMean}
\small
	\begin{aligned}
    & \mathbf{C}_e^k = \sigma \left(\frac{1}{|\mathcal{K}(e)|}\left( \sum_{w\in  \mathcal{K}(e)}{\mathbf{E}_w^{k-1}} \right)W_1^k+ b_{1}^k\right), \\
	\end{aligned}
\end{equation}
where $W_1^k{\in}\mathbb{R}^{d\times d}$ and $b_{1}^k{\in}\mathbb{R}^d$ denote the transformation matrix and bias to be learned, and we choose \texttt{LeakyReLU} as the nonlinear activation function.

% \textbf{Relation-aware User Modeling}
% \chen{remove it and re-organize the section-name-structure, Zirui: DONE}

The triplet social relation encoded by hyperedges conveys the inhomogenous social influence between users. 
Since two users can be connected by multiple hyperedges, to well model the social relation, we need to aggregate the effect of different hyperedges. Therefore we design the following hypergraph convolutional layer to obtain the representation of social relations. 
% User modeling is divided into two parts to make the propagation process relation-aware. Firstly, relationship between any two connected users would be represented as a vector based on the hyperedges' representation. Then, users' latent preferences would be propagated through networks gated by hyperedges and relationships.
% To make our model relation-aware, we first use hyperedges' representation to deviate representation of relationship for any two connected users, then 
%  Relationship modeling is inspired by the heuristic idea that the relationship between two linked users are shaped by all the interactions they shared. Item modeling is basically following the framework of \textsl{Graph Convolution Network} (GCN), where the messages were passed from both the hyperedges and users who had purchased them. The massages passed to users are made up of three parts: messages from adjacent hyperedges and messages from social friends.
\subsubsection{Social Relation Representation Module}
The existing work of modeling complex relations among nodes~\cite{schlichtkrull2018modeling} tries to learn a function $f$ defined on the power set of the node-set, of which the output would describe the relation between the input nodes. However, this modeling method, which roughly takes the node-set as input, ignores the internal structure of the graph.

To address it, we build a vector $R_t$ to represent the social relation for two users, $i_1$ and user $i_2$, who are friends.
% In the framework of graph convolution network, we can still use the concept of aggregator to illustrate the idea of our relation modeling method.
To obtain the representation, we design the graph convolutional layers as follows,
\begin{equation}\label{equation:relationModeling}
\small
	\begin{aligned}
    & \mathbf{R}_t^k = \sigma \left({\textit{aggregate}} \left(\mathbf{C}_e^k | e\in  \mathcal{N}(i_1, i_2)\right)\right),  t = \eta(i_1, i_2),\\
	\end{aligned}
\end{equation}
where $\mathcal{N}(i_1, i_2)$ denotes the set of hyperedges connecting to both user $i_1$ and user $i_2$, and $\eta(i_1, i_2)$ is the mapping function that outputs the index $i_1$-$i_2$ relation for embedding lookup from $\mathbf{R}$.

We would like to emphasize the advantages of this social-relation modeling method.
We obtain a low dimensional vector representation of the two social-connected users in the process of message passing from the triplet social relations. Since the triplet social relation, such as item-sharing or co-purchase behaviors, reflects the inhomogenous social influence, this process of message passing can help adaptively aggregate all triplet social relations between two users.
% Therefore, we do not only take the characteristics of the connected users and item, or namely the embedding of the nodes, but also the structure of the graph into consideration. In particular, if two users often share items to each other, or they have much more common friends than the average level, they relationship also get more message from the graph.

In our design method, we implement the aggregation in Eqn (\ref{equation:relationModeling})
with the graph convolutional networks, propagating the message from hyperedge to relation embedding. Then Eqn (\ref{equation:relationModeling}) can be re-formulated as follows,
\begin{equation}\label{equation:relationMean}
\small
	\begin{aligned}
    & \mathbf{R}_t^k = \sigma \left(\frac{1}{|\mathcal{N}(i_1, i_2)|}\left( \sum_{e\in  \mathcal{N}(i_1, i_2)}{\mathbf{C}_e^k}\right)W_2^k+ b_{2}^k\right), t = \eta(i_1, i_2)\\
	\end{aligned}
\end{equation}

The empirical evidence in Section~\ref{sec::exp} shows that this aggregator works well on real-world datasets.
% To summarize, via this module, we obtain the representation of user-user social relation.

\subsubsection{User Modeling Module}
In social recommendation, a user's characteristics are partly built from his/her friends' characteristics.
This commonly-known \textit{social-trust} effect, is widely considered in existing social recommendation works~\cite{jamali2010matrix, wu2019neural}.
However, these works are limited to only using scalar weight, which means the strength of social-relation describing the social trusts.
In our method, as we have obtained the social relations' representations, it becomes possible to further enhance the modeling of the social trusts.

The key idea here is to additionally use the social relation representation calculated in equation (\ref{equation:relationModeling}) to model the social relation between two users in a fine-grained manner.

We can deploy a message passing-based aggregation process formulated as follows,
\begin{equation}\label{equation:userModeling}
\small
	\begin{aligned}
    & \mathbf{P}_i^k = \mathbf{P}_i^{k-1} + \\
    &\underbrace{\sigma \left({\textit{aggregate}} \left(\mathbf{C}_e^k | e\in  \mathcal{Z}(i)\right)\right)}_{\text{messages from hyperedges}} + \underbrace{\sigma \left({\textit{aggregate}} \left(\mathbf{P}_w^{k-1} | w\in  \mathcal{N}(i)\right)\right)}_{\text{messages from social network}},\\
	\end{aligned}
\end{equation}
where the messages from hyperedges help address the limitations of existing works that can only aggregate messages from social network. Here $\mathcal{Z}(i)$ denotes the set of hyperedges connected to user $i$. 
As shown above, we also add a self-loop operation for the user embedding, of which the embedding of the previous layer would be directly added to the embedding of the next layer. This would help prevent the vanishing or exploding gradients problem.

Messages from hyperedges aim to capture the latent signals of inhomogeneous social effects, while messages from social network mainly concentrate on social homophily modeling. In other words, the first term varies between different items while the second term measures how similar two users are in general.

In our experiment, we implement the aggregation in Eqn (\ref{equation:userModeling}) via
an attentive graph convolutional layer.
Specifically, multi-layer perceptron is adopted to generate the attention weight of user $w$ with respect to $P_i^k$.
Therefore the equation (\ref{equation:userModeling}) can be reformulated as follows,
\begin{equation}\label{equation:userATT}
\small
	\begin{aligned}
	&\mathbf{P}_i^k = \mathbf{P}_i^{k-1} + \\
    &\sigma \left(\frac{1}{|\mathcal{Z}(i)|}\left( \sum_{e\in  \mathcal{Z}(i)}{\mathbf{C}_e^k} \right)W_3^k + b_{3}^k \right)  + \sigma \left(\sum_{w\in \mathcal{N}(i)}{\alpha _{wi}^k\mathbf{P}_w^k}\right),\\
    &\alpha _{wi}^k  =\text{MLP}\left(\mathbf{R}_{\eta(w,i)}\right),
	\end{aligned}
\end{equation}
where $\text{MLP}$ is used to learn the attention weights $\alpha _{wi}$ for all the friends of user $i$. 
In our experiment, we find that the recommendation performance is not sensitive to MLP's structure.

\subsubsection{Item Modeling Module}
As for the item, it can also take full use of the hyperedges' representations for capturing item features.
In fact, the internal features of a specific item are indeed shaped by the characteristics of the triplet social relations that involve this item. For example, the item-sharing behaviors in social network are highly related to the item itself, besides the two users.

Therefore, we can propose a propagation module to aggregate the hyperedges' representations formulated as follows,

\begin{equation}\label{equation:itemModeling}
\small
	\begin{aligned}
    & \mathbf{Q}_j^k = \mathbf{Q}_j^{k-1} + \sigma \left({\textit{aggregate}} \left(\mathbf{C}_e^k | e\in  \mathcal{Z}(j)\right)\right),\\
	\end{aligned}
\end{equation}
where $\mathcal{Z}(j)$ denotes the set of hyperedges connected to item $j$. Similarly to Eqn (\ref{equation:userModeling}), there is also a self-loop term $\mathbf{P}_i^{k-1}$.

We implement the aggreation via the plain GCN aggregator and then Eqn~(\ref{equation:itemModeling}) can be reformulated as follows,
\begin{equation}\label{equation:itemMean}
\small
	\begin{aligned}
    & \mathbf{Q}_j^k = \mathbf{Q}_j^{k-1} + \sigma \left(\frac{1}{|\mathcal{Z}(j)|}\left( \sum_{e\in  \mathcal{Z}(j)}{\mathbf{C}_e^k}\right)W_4^k + b_{4}^k\right),\\
	\end{aligned}
\end{equation}
where $W_4^k$ and $b_{4}$ are learnable transformation matrix and bias vector.
Note that we normalize the output embedding of each layer to ensure the numerical stability.
It is worth-mentioning that although some more complicated aggregator such as attention-base aggregator can also be adopted, this GCN aggregator has shown promising experimental results. Therefore we leave the exploration as the future work.

\subsection{Prediction and Optimization}
\subsubsection{Prediction Layer}
After the propagation of $L$ layers, we can obtain $L$ different embeddings of both users and items $\{E^0,\cdots,E^L\}$. 
Follow existing works~\cite{wang2019neural}, we concatenate the output of all layers to generate the final representation of users and items as follows,
\[\mathbf{E}^* = \mathbf{E}^0 || \cdots || \mathbf{E}^L = [\mathbf{P}^*, \mathbf{Q}^*].\] 
We then choose the simple yet effective inner product as our prediction function as follows,
\begin{equation}
   {r_{ij}} = \mathbf{P}_i^* \otimes  \mathbf{Q}_j^*.
\end{equation}

\subsubsection{Model Optimization}
To optimize our model parameters, we adopt the BPR loss~\cite{rendle2009bpr} that is widely used in implicit recommender systems~\cite{wang2019neural,he2017neural,wu2019neural,DiffNet++} as follows,
\begin{equation}\label{equ:loss}
	Loss = \sum_{(i,j_1,j_2)\in \mathcal{O}}{-\ln\sigma({r}_{ij_1}-{r}_{ij_2})+\lambda{\left\|\Theta\right\|}^{2}_{2}},
\end{equation}
where $\mathcal{O} = \{(i,j_1,j_2)|(i,j_1){\in}\mathcal{Y}, (i,j_2){\in}\mathcal{Y}^- \}$ denotes the pairwise training data with negative sampling, and $\mathcal{Y}$ and $\mathcal{Y}^-$ denote the observed and sampled unobserved user-item interaction set, respectively.

Thanks to the Automatic Differentiation framework like TensorFlow ~\cite{abadi2016tensorflow} and PyTorch ~\cite{paszke2019pytorch},
we are freed from the complex gradient computation process. Therefore, we omit the computation of gradient with respect to parameters.

\subsection{Model Size and Time Complexity}
We would like to first discuss the model size and time complexity of our proposed model.
\vspace{0.2cm}

\noindent \textbf{Model Size.}
It is worth mentioning that SHGCN is a fairly lightweight model, although we introduce four embedding matrices $(\mathbf{C}^k,\mathbf{R}^k,\mathbf{P}^k,\mathbf{Q}^k)$ at each hypergraph convolutional layer ($k$ denotes the depth). 
The trainable parameters of SHGCN contain three parts, embeddings matrices of users and items, parameters of MLP, and parameters of four linear transformations in each layer. 
For the first part, only the full embedding matrix of the $0$-th layer $\mathbf{E}^0{=}\mathbf{E}{=}[\mathbf{P}, \mathbf{Q}]{\in}\mathbb{R}^{(M+N){\times}d}$ is trainable. 
For the second part, we employ a two-layer MLP in our experiment, so we only have $2L{\cdot}d^2$ extra parameters. 
For the last part, each linear transformation operation introduces parameters with size $d(d+1)$ and in total it introduces $4L{\cdot}d(d+1)$-size parameters. 

Considering the fact that $\min(M,N){\gg}\max(L,d)$, we can claim our SHGCN is as lightweight as MF~\cite{koren2009matrix} --- one of the most concise embedding-based recommender model (only has the first part of parameters). 
Take our experimented Beibei dataset as an example. 
There are 150K users and 30K users in Beibei. If we set the embedding size as 32 and use 3 propagation layers, MF has 5.76 million parameters while our SHGCN introduces only 18K additional parameters. In other words, our model uses only around 3\% more parameters compared with MF.

\vspace{0.2cm}

\noindent \textbf{Time Complexity}
In the whole process of training, the most time-consuming part of our model is four aggregate operations in the hypergraph convolutional layer:
hyperedge representation module, social-relation representation module, user modeling module, and item modeling module. The time complexity of these four modules are  $O(|\mathcal{E}|d)$,  $O(|\mathcal{T}|d)$,  $O((|\mathcal{T}|+|\mathcal{E}|)d)$, and $O(|\mathcal{E}|d)$, respectively. Here $|\mathcal{E}|$ and $|\mathcal{T}|$ denote number of hyperedges and social relations. As we can observe, our model has a linear time complexity with respect to the scale of the dataset. 

Empirically, hyperedges are very sparse, thus the additional time expense is relatively small. In our experiment,  MF and our SHGCN cost around $14s$ and $20s$ per epoch on Beibei dataset, respectively, under the same embedding size, when training on a Tesla V100 GPU. In contrast, a complicated hypergraph convolution network-based method, MHCN~\cite{yu2021selfsupervised}, costs around $60s$, which is twice more time-consuming than our model.

\vspace{0.2cm}
\subsection{Open Discussions}
The proposed framework is blessed with the natural advantage of hyperedge, and thus it can be adapted to hyperedges with arbitrary degrees, which guarantees the prospect of extensive application scenarios. 
Even though we only evaluate on tackling triple social relation, the proposed method, as presented by Eqn (\ref{equation:hyperedgeModeling})-(\ref{equation:itemMean}) remains the same when deploying our model on much more complex interactions.

Actually, some well-studied problems such as product-list recommendation can also be regarded as a problem of handling one-to-many relations.
We argue that the framework of SHGCN can also help in those scenarios, and we leave this for future studies.

In addition, in the process of propagation, two intermediate embeddings are derived, namely the embedding of hyperedges and social relations. 
The derived embedding of hyperedges and social relations can also be used in downstream tasks like link prediction or classification, which will be an interesting research problem.

\section{Experiments}\label{sec::exp}
In this section, we conduct experiments on two real-world datasets to evaluate the proposed model. We aim to answer the following three research questions.
\begin{itemize}[leftmargin=*]
    \item \textbf{RQ1}: How does our method perform comparing to the state-of-the-art models? Does introducing the explicit modeling of triple social relations via hyperedge improve the recommendation performance?
    \item \textbf{RQ2}: Can our proposed model address the data sparsity issue? In other words, can our model still steadily outperform baselines for users with fewer interactions?
    \item \textbf{RQ3}: How do the hyper-parameters affect our model's recommendation performance? In other words, does our model need lots of effort in tuning hyper-parameters to achieve good recommendation performance?
\end{itemize}

\subsection{Experimental Settings}
\subsubsection{Dataset}
We conduct experiments on two datasets with different scales collected from real-world applications. 
% It should be noted that interactions between multiple users or items are needed, so no public dataset is suitable for our problems. These two datasets are variants in scale and application scenarios. 
The statistics of two utilized datasets are reported in Table \ref{tab::DataStat}. 
% For each dataset, we do some simple filtering to prevent the potential data breaches by avoiding entries of sharing interaction where the shared item of a specific user coincides with the positive sample in the test set.
% The dataset along with codes are provided in this anonymous link\footnote{XXX}.

\begin{itemize}[leftmargin=*]
    \item  \textbf{Beidian.} This dataset is
    released by~\cite{lin2019cross}, collected from Beidian\footnote{https://www.beidian.com}, which is an e-commerce platform that supports users sharing products' URL links in social network. This dataset contains two parts of data, purchase-behavior logs and item-sharing logs. Therefore the triplet social relation involves a user sharing the item, the shared item, the user receiving the shared item.
    % In this dataset, two users in one item-sharing record are treated equally, which means the hyperedges that model the sharing behaviors are non-directional.
\item  \textbf{Beibei.} This dataset is released by~\cite{zhang2021group}, collected from Beibei\footnote{https://www.beibei.com},
which is the largest e-commerce platform for maternal and infant products in China. In this platform, similar to Pinduoduo.com, group-buying is the most popular manner for users to purchase products. A user can launch a group-buying and invite his/her friends to join via sending URL links in social network.
Each entry in this dataset represents a group buying behavior, and we select the group buying logs having two users and one item since most group-buying behaviors are at the two-user size.
Therefore the triplet social relation involves two users in the group and the purchased item.

% to keep pace with the Beidian. 
% The interactions of purchase are considered to take place between every user in the group and the related item.
\end{itemize}

These two datasets to evaluate our model reflect two of those representative real-world scenarios where we can collect the data reveals inhomogeneous social relation, item-sharing and group-buying. Since these scenarios have achieved remarkable commercial success such as Pinduoduo.com, our studied problem has vast and valuable applications.
% \chen{weaken the particularity of our datasets and emphasize that our method has wide applications}

\begin{table}[]
\centering
\caption{Statistics of the datasets.}
\label{tab::DataStat}
\begin{tabular}{c|c|c|c|c}
\hline
Dataset & \#User  & \#Item & \#U-U-I & \#U-I     \\ \hline
Beidian & 3,773   & 4,544  & 9,358   & 39,252    \\ \hline
Beibei  & 149,361 & 30,486 & 522,264 & 1,089,266 \\ \hline
\end{tabular}
\end{table}

\subsubsection{Evaluation Protocals}
Following existing works~\cite{he2017neural}, we apply the \textit{leave-one-out} evaluation protocol to evaluate the performance of our model, and datasets are divided into the training set, validation set and the testing set.

To evaluate the overall performance of all models, we adopt the two most commonly used metrics, \textit{Recall} and \textit{NDCG}, in recommender systems, defined as follows,
\begin{itemize}[leftmargin=*]
	\item \textbf{Recall@K}: Recall in the recommender systems indicates the probability that the true positive item is present in the top-K recommended list in a statistical sense.
	\item \textbf{NDCG@K}: \textit{Normalized Discounted Cumulative Gain} (NDCG) is an enhancement to Recall by taking the ranking location into consideration instead of merely counting whether the true positive item is hit.
\end{itemize}
Specially, NDCG is equal to Recall when the metrics are evaluated in Top-$1$ list.

\begin{table*}[t]
\renewcommand\arraystretch{1.4}
\centering
\caption{Overall Performance on the Beidian dataset (all $p$-value\textless 0.01).\\Note that Recall has the same value as NDCG when K=1.}
\label{tab::overall-Beidian}
\begin{tabular}{c|c|c|ccccccc}
\hline
Dataset                   & Category                                 & Method               & NDCG@1          & Recall@3        & NDCG@3          & Recall@5        & NDCG@5          & Recall@10       & NDCG@10  \\ \hline
\multirow{11}{*}{Beidian} & \multirow{4}{*}{Collaborative Filtering} & MF                   & 0.1724          & 0.3008          & 0.2467          & 0.3691          & 0.2746          & 0.4757          & 0.3093   \\
                          &                                          & GraphSage-BG         & 0.1555          & 0.3025          & 0.2405          & 0.3716          & 0.2688          & 0.4984          & 0.3096   \\
                          &                                          & NGCF-BG              & {\ul 0.1762}    & {\ul 0.3123}    & {\ul 0.2552}    & {\ul 0.3913}    & {\ul 0.2877}    & 0.5041          & {\ul 0.3239}   \\
                          &                                          & LightGCN             & 0.1519          & 0.2781          & 0.2241          & 0.3557          & 0.2560          & 0.4754          & 0.2945   \\ \cline{2-10} 
                          & \multirow{5}{*}{Social   Recommendation} & SocialMF             & 0.1637          & 0.3052          & 0.2454          & 0.3719          & 0.2729          & 0.4828          & 0.3089   \\
                          &                                          & GraphSage-EG         & 0.1541          & 0.2923          & 0.2331          & 0.3751          & 0.2671          & 0.4926          & 0.3050   \\
                          &                                          & NGCF-EG              & 0.1667          & 0.3008          & 0.2447          & 0.3776          & 0.2763          & 0.5005          & 0.3161   \\
                          &                                          & Diffnet              & 0.1724          & 0.3087          & 0.2514          & 0.3877          & 0.2841          & 0.5003          & 0.3204   \\
                          &                                          & MHCN                 & 0.1727          & 0.3085          & 0.2506          & 0.3836          & 0.2815          & 0.4910          & 0.3163   \\\cline{2-10} 
                          & \multirow{2}{*}{Adapted Methods}         & GraphSage-CG         & 0.1626          & 0.3011          & 0.2427          & 0.3732          & 0.2725          & 0.4948          & 0.3118   \\
                          &                                          & NGCF-CG              & 0.1708          & {\ul 0.3123}    & 0.2524          & 0.3899          & 0.2842          & {\ul 0.5098}    & 0.3231   \\ \cline{2-10} 
                          & \textbf{Proposed Method}                 & \textbf{SHGCN}       & \textbf{0.1814} & \textbf{0.3363} & \textbf{0.2715} & \textbf{0.4153} & \textbf{0.3039} & \textbf{0.5309} & \textbf{0.3414} \\ \cline{2-10} 
                          & \multicolumn{1}{l|}{}                    & \textbf{Improvement} & 2.95\%          & 7.68\%          & 6.39\%          & 6.13\%          & 5.63\%          & 4.14\%          & 5.40\%   \\ \hline 
\end{tabular}
\end{table*}

\begin{table*}[t]
\renewcommand\arraystretch{1.4}
\centering
\caption{Overall Performance on the Beibei dataset (CG-version methods in Beibei dataset is equivalent to their EG counterparts as user-item interactions is actually extracted from the group-buying triplets; all $p$-value\textless 0.01).}
\label{tab::overall-Beibei}
\begin{tabular}{c|c|c|ccccccc}
\hline
Dataset                 & Category                                 & Method               & NDCG@1            & Recall@3          & NDCG@3          & Recall@5       & NDCG@5       & Recall@10     & NDCG@10     \\ \hline
\multirow{9}{*}{Beibei} & \multirow{4}{*}{Collaborative Filtering} & MF                   & 0.1211            & 0.2479            & 0.1942          & 0.3264         & 0.2265       & 0.4489        & 0.2660      \\
                        &                                          & GraphSage-BG         & 0.1250            & 0.2558            & 0.2004          & 0.3350         & 0.2329       & 0.4612        & 0.2737      \\
                        &                                          & NGCF-BG              & 0.1326            & 0.2642            & 0.2083          & 0.3445         & 0.2412       & 0.4681        & 0.2812      \\ 
                        &                                          & LightGCN             & 0.1195            & 0.2503            & 0.1948          & 0.3324         & 0.2285       & 0.4615        & 0.2701      \\ \cline{2-10} 
                        & \multirow{5}{*}{Social Recommendation}   & SocialMF             & 0.1235            & 0.2509            & 0.1968          & 0.3297         & 0.2292       & 0.4531        & 0.2690      \\
                        &                                          & GraphSage-EG         & {\ul 0.1327}      & {\ul 0.2710}      & {\ul 0.2124}    & 0.3525         & {\ul 0.2459} & 0.4789        & 0.2867      \\
                        &                                          & NGCF-EG              & 0.1263            & 0.2626            & 0.2046          & 0.3469         & 0.2393       & 0.4756        & 0.2809      \\
                        &                                          & Diffnet              & 0.1311            & 0.2685            & 0.2102          & {\ul 0.3528}   & 0.2448       & {\ul 0.4811}  & {\ul 0.2863}  \\
                        &                                          & MHCN                 & 0.1093            & 0.2274            & 0.1772          & 0.3020         & 0.2078       & 0.4229        & 0.2468       \\ \cline{2-10} 
                        & \textbf{Proposed Method}                 & \textbf{SHGCN}       & \textbf{0.1503}   & \textbf{0.2882}   & \textbf{0.2298} & \textbf{0.3680} & \textbf{0.2626}  & \textbf{0.4916}  & \textbf{0.3025}     \\ \cline{2-10} 
                        & \multicolumn{1}{l|}{}                    & \textbf{Improvement} & 13.26\% & 6.35\% & 8.19\% & 4.31\% & 6.79\% & 2.18\% & 5.51\% \\ \hline
\end{tabular}
\end{table*}

\subsubsection{Baselines}
We adopt the following representative and state-of-the-art methods as baselines for performance comparison. These methods are generally divided into three categories, including collaborative filtering methods, social recommendation methods and adapted methods. 
% \chen{please divide them into groups similar as Table III, Zirui: DONE}.

Collaborative filtering methods refer to those models that can only utilize user-item interaction data. The collaborative filtering methods we choose are introduced as follows:

\begin{itemize}[leftmargin=*]
	\item \textbf{MF~\cite{koren2009matrix}. }This is a competitive matrix factorization method, which is the cornerstone of most state-of-art recommendation algorithms. 
	\item \textbf{SocialMF~\cite{jamali2010matrix}. }
	This is a famous matrix factorization-based social recommendation model. This method optimizes a pairwise loss, introducing an additional social regularization term to model social relation, calculated as the distance between a user and his/her friends' weighted sum.
	\item \textbf{GraphSage-BG~\cite{hamilton2017inductive}: }GraphSAGE~\cite{hamilton2017inductive} is one of the most widely used GCN framework which enriches node embedding with its neighbors’ information by embedding propagation and aggregation. We deploy GraphSAGE to the user-item \underline{B}ipartite \underline{G}raph and name it as GraphSage-BG. Therefore it is a collaborative filtering method. 
	\item \textbf{NGCF-BG~\cite{wang2019neural}:} NGCF is one of the state-of-the-art GCN-based models which exploits the user-item graph structure by modeling high-order connectivity and injecting the collaborative signal through propagation. We deploy NGCF to the user-item bipartite graph and name it NGCF-BG.
\end{itemize}

Social recommendation methods can leverage the social network as side information to infer user preferences. In our experiment, we construct the social network by treating two users as connected as long as they ever appear in a same hyperedge. The compared social recommendation methods are as follows:

\begin{itemize}[leftmargin=*]
    \item \textbf{GraphSage-EG:} We further merge the social network into the user-item bipartite graph to get an \underline{E}xtended \underline{G}raph. We then conduct the propagation and aggregation of GraphSage on the extended graph and name this method as GraphSage-EG.   
	\item \textbf{NGCF-EG:} NGCF-EG is constructed in a similar way to GraphSage-EG. The propagation on the extended graph makes NGCF aware of the social network structure even though it is originally designed to exploit the user-item graph only. 
	\item \textbf{Diffnet~\cite{wu2019neural}: }Diffnet is a recently-advanced GCN-based method for social recommendation. It adopts the graph convolutional layers in the social graph and achieves state-of-art performance due to its ability to simulate the recursive social diffusion process through a layer-wise influence propagation structure.
	\item \textbf{MHCN~\cite{yu2021selfsupervised}: }\textcolor{black}{MHCN is a most recent multi-channel hypergraph convolutional network-based method which works on multiple motif-induced hypergraphs. An InfoNCE-like\cite{oord2019representation} loss is added to MHCN's learning objective to maximize the hierarchical mutual information.}
\end{itemize}

Some classic and commonly used baselines cannot even tackle hyperedges as input. Thus, we adapt the following methods to make them more compatible with our problems.

\begin{itemize}[leftmargin=*]
	\item \textbf{GraphSage-CG:} We convert the hyperedge structure to a undirected \underline{C}omplete \underline{G}raph structure and deploy GraphSage on this graph. More specifically, for each hyperedge, we connect every pair of nodes in this hyperedge with an ordinary edge. We name this version GraphSage-CG.
	\item \textbf{NGCF-CG:} We adapt NGCF in the same practice with GraphSage-CG to enable it to tackle the hyperedge input. We name this version NGCF-CG.
\end{itemize}

\subsubsection{Hyper-parameter Settings}\label{subsectionsetting}
All models mentioned above, including baselines and our SHGCN are trained with the BPR loss~\cite{rendle2009bpr}, which is commonly used in recommender system. We randomly select eight items for one entry in the train set as negative samples to train all the model. We adopt sampled metrics to evaluate the performance of all methods following~\cite{he2017neural,wu2019neural}, where we randomly select 100 items for one entry in the test set.
Following existing works~\cite{GraphRec,wu2019neural}, we deploy Adam optimizer with the 4096-size mini-batch and fit the embedding size $d=32$ for all aforementioned methods. The learning rate is tuned in \{3e-4, 1e-3, 3e-3\} and $L_2$ regularization term is searched in \{1e-8, 1e-7, 1e-6, 1e-5, 1e-4, 1e-3\}. We fixed the message dropout and node dropout to 0 when it comes to GCN-based methods, as the methods preventing over-fitting is not our main concentration. We set the number of GCN layers to three, which has been demonstrated good choice by existing works~\cite{wang2019neural,wang2019kgat}.

\subsection{Overall Performance (RQ1)}
We first present the overall performance on two utilized datasets in Table~\ref{tab::overall-Beidian} and Table~\ref{tab::overall-Beibei}, respectively.
For the two top-K metrics, we set top-K to \{1, 3, 5, 10\}, a widely used range in existing works~\cite{wu2019neural}.
From these results, we have the following observations.
\begin{itemize}[leftmargin=*]
    \item \textbf{\textcolor{black}{SHGCN outperforms all the baselines significantly in two real-world datasets}.
    }
    As we can observe from Table \ref{tab::overall-Beidian} and Table~\ref{tab::overall-Beibei}, our SHGCN outperforms all the baselines significantly on all Recall@K and NDCG@K metrics. \textcolor{black}{We arbitrarily select five random seeds, repeat the training process, and report the averaged results. The p-values of independent two-sample t-tests are smaller than 0.01 for all metrics showed above, demonstrating the performance improvement is significant and steady. From these results, SHGCN's effectiveness is well validated.
    } 
    \item \textbf{Existing graph-based methods fail to model high-dimensional social relations.} We can observe from Table~\ref{tab::overall-Beibei} and~\ref{tab::overall-Beidian} that directly incorporating the social interactions to the graph does not make GraphSage or NGCF performs better due to they fail to capture the latent signals passed by the inhomogeneous social relations.
    In contrast, SHGCN explicitly extracts messages from triple social relations and uses social relations' representation to shape the user embedding.
    From this point of view, our model succeeds in utilizing the triple social relations by effective hypergraph and graph convolutonal layers.
    \item \textbf{Traditional social recommendation baselines fail to utilize triple social relations.}
    Traditional social recommendation methods utilize the social network structure by minimizing the distance of connected users in the embedding space. SocialMF directly adds the distance term to its learning objective, while GCN-based methods implicitly achieve this purpose through the propagation on the graph. Hypergraph-based baseline MHCN highly depends on manually designed motifs and channels based on human expertise, and thus it lacks the ability to generalize when facing new datasets whose underlying semantics are different. As we can observe from Table~\ref{tab::overall-Beibei} and Table~\ref{tab::overall-Beidian}, MHCN fails catastrophically on Beibei dataset.
    Our SHGCN is designed in a more elegant way as it does not depend on any manually designed motifs or propagation paths. Message passing patterns are learned automatically in the process of hypergraph convolution. Therefore, we can expect a more consistent performance on different datasets, and empirical results also turn out this way.
\end{itemize}

\begin{figure*}[!t]
	\begin{center}
		\mbox{\subfloat{\includegraphics[scale=0.65]{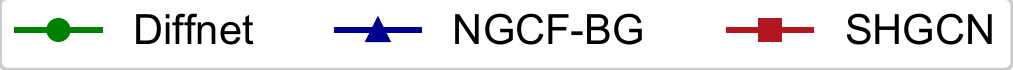}}}
	\end{center}
	\vspace{-0.8cm}
\end{figure*}
\begin{figure*}[!t]
	\begin{center}
		\mbox{
			% original: width=8.5
			\subfloat{\includegraphics[width=0.25\linewidth]{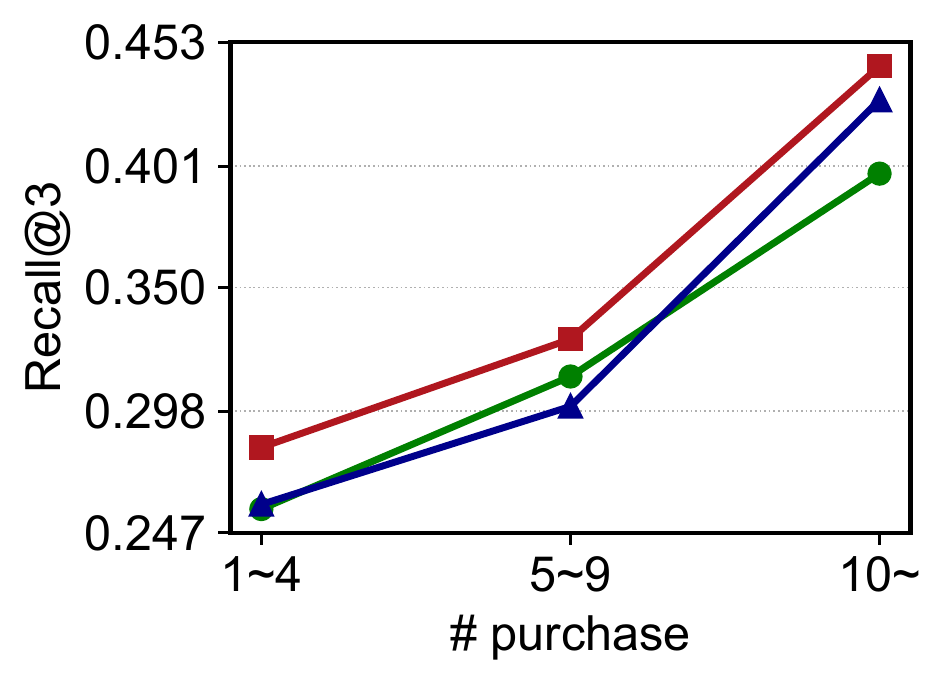}}
			\subfloat{\includegraphics[width=0.25\linewidth]{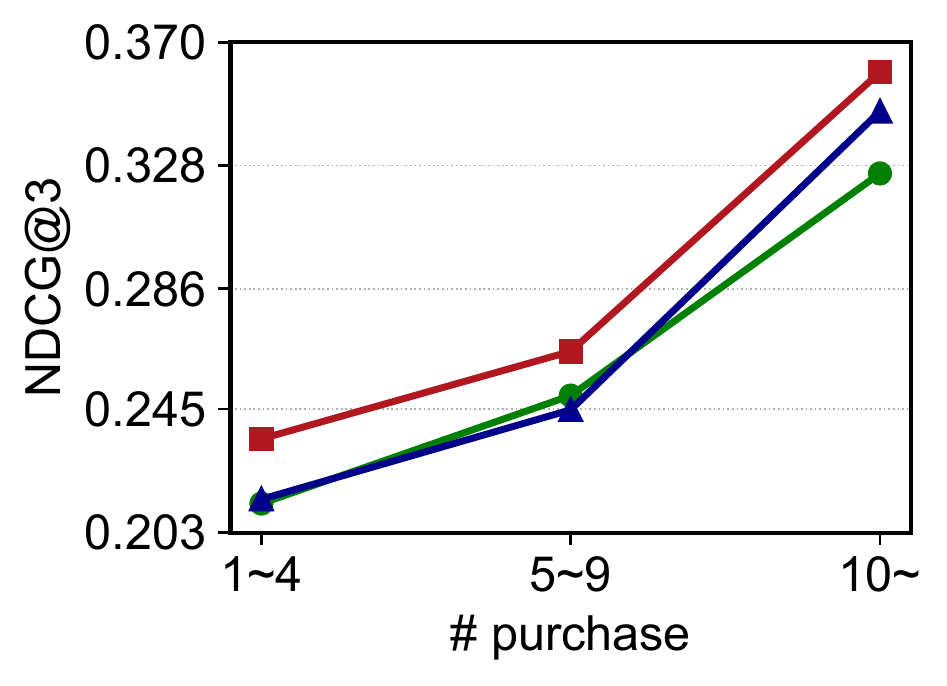}}
			\subfloat{\includegraphics[width=0.25\linewidth]{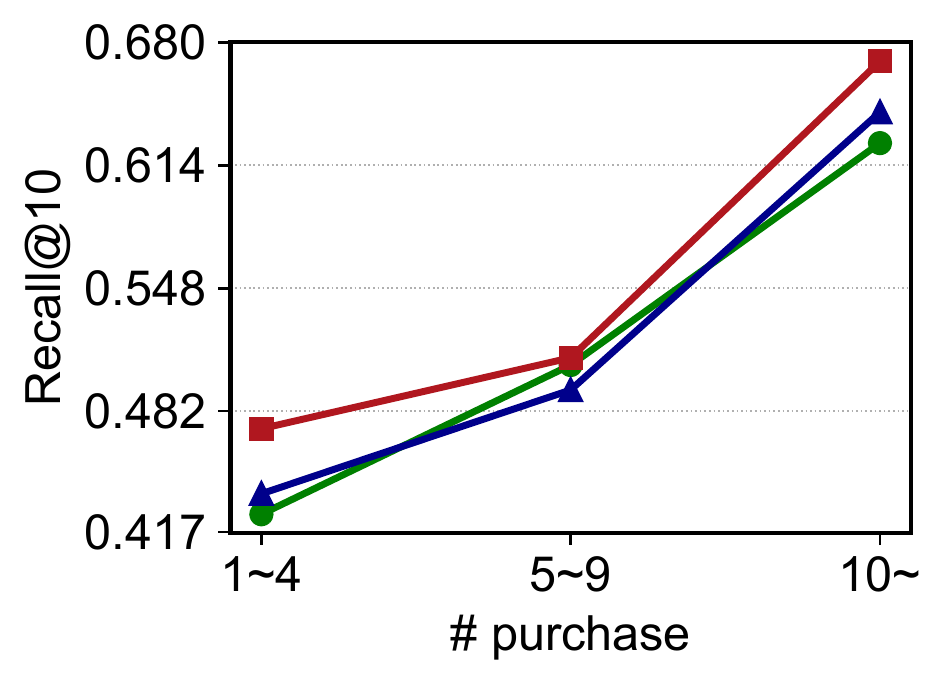}}
			\subfloat{\includegraphics[width=0.25\linewidth]{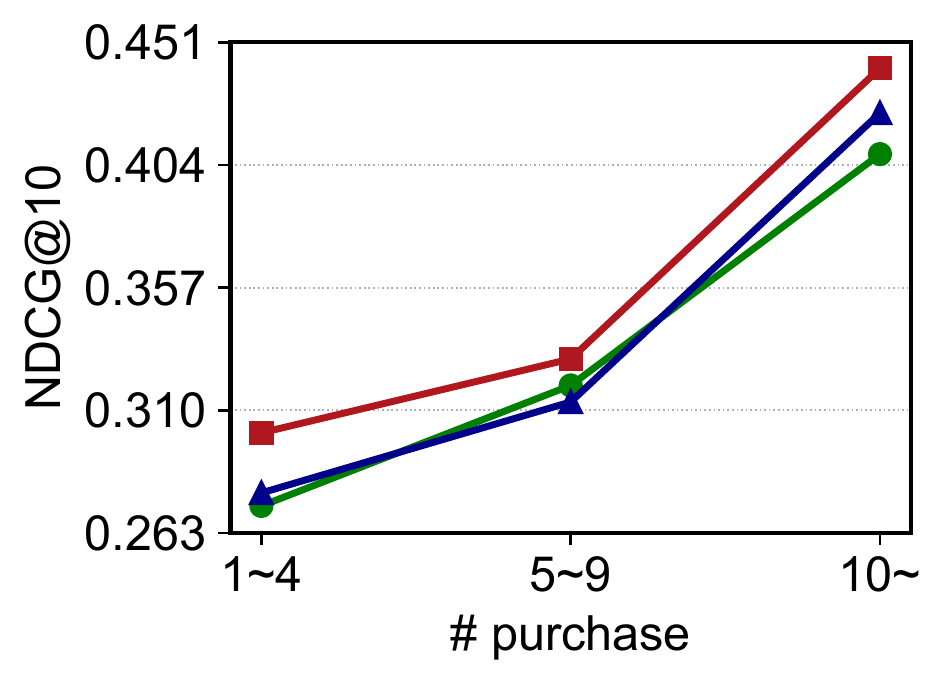}}
		}
	\end{center}
	\caption{Recommendation performance for users with different number of interactions on the Beidian dataset}\label{fig::rq2::beidian}
\end{figure*}

\begin{figure*}[!t]
	\begin{center}
		\mbox{
			% original: width=8.5
			\subfloat{\includegraphics[width=0.25\linewidth]{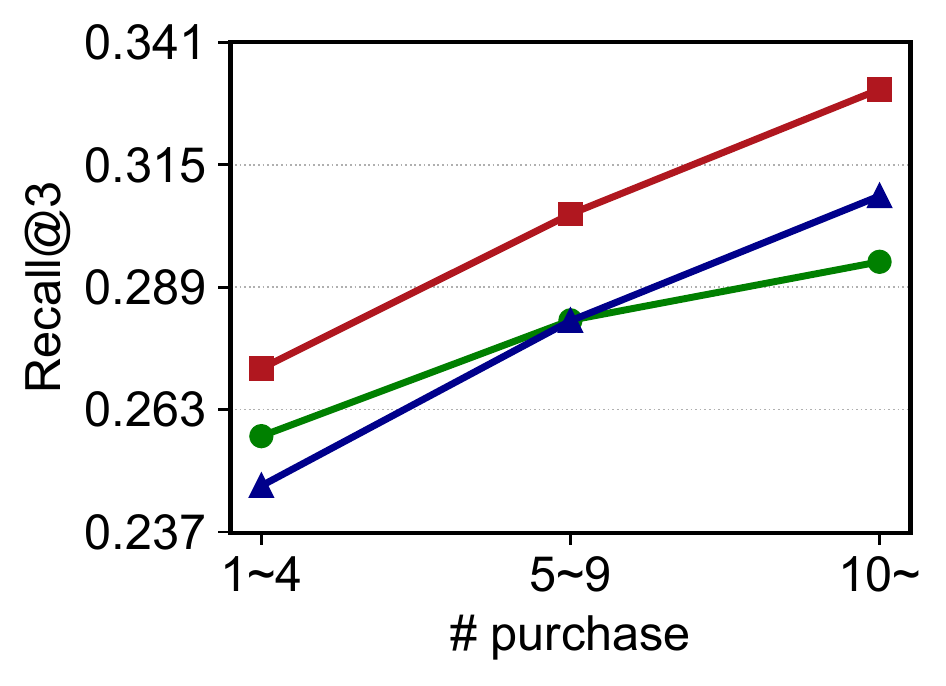}}
			\subfloat{\includegraphics[width=0.25\linewidth]{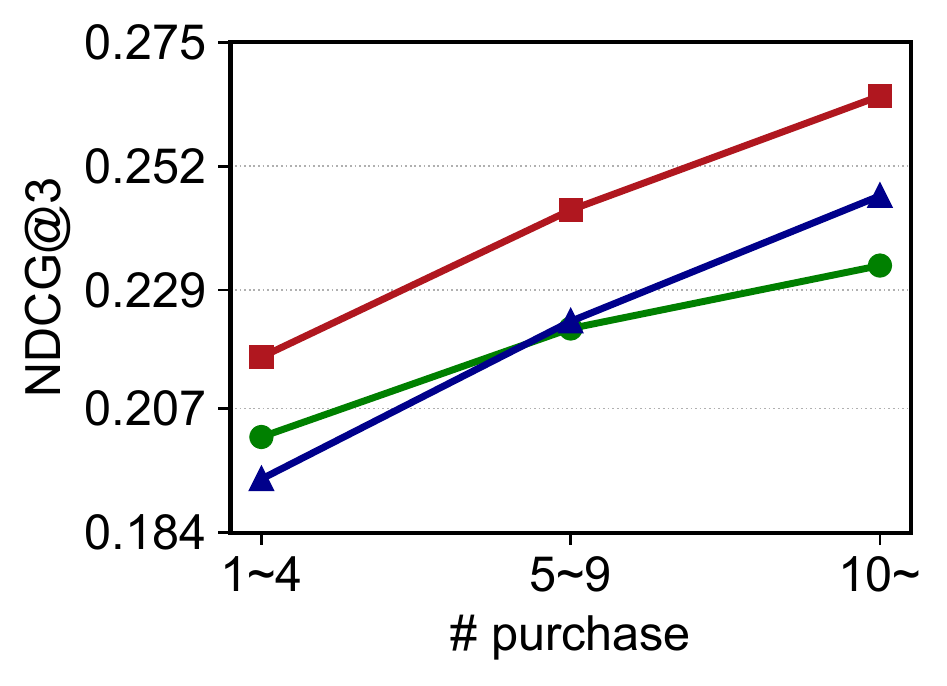}}
			\subfloat{\includegraphics[width=0.25\linewidth]{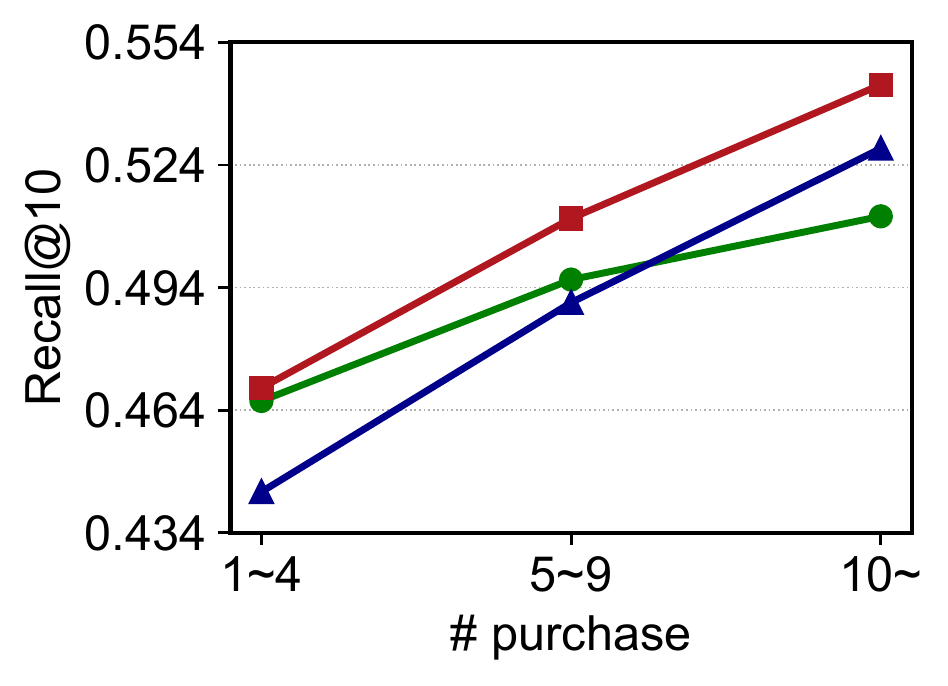}}
			\subfloat{\includegraphics[width=0.25\linewidth]{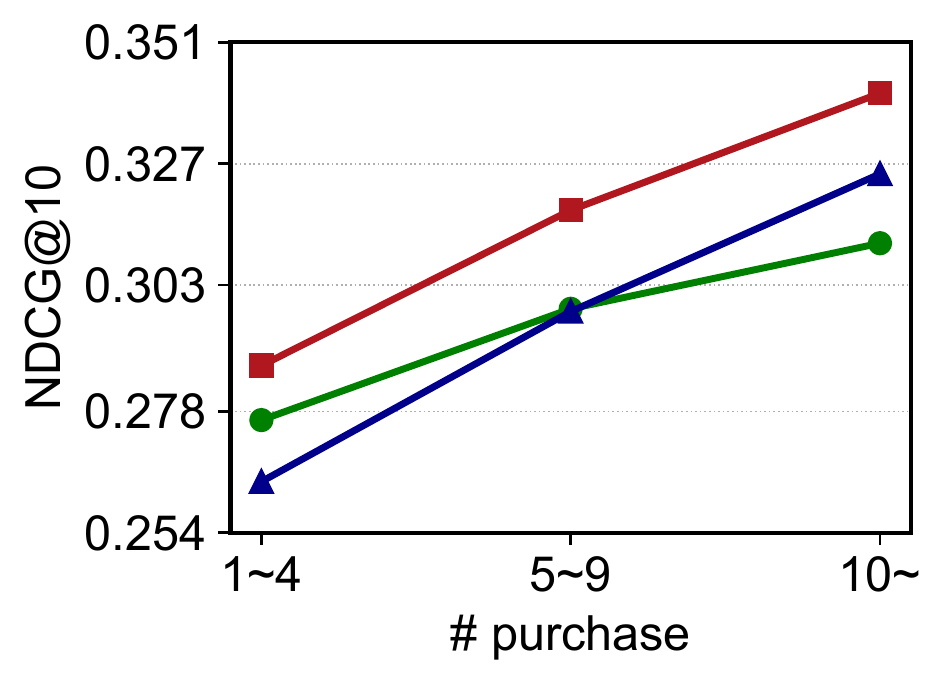}}
		}
	\end{center}
	\caption{Recommendation performance for users with different number of interactions on the Beibei dataset}\label{fig::rq2::beibei}
\end{figure*}

\subsection{Data Sparsity Issue (RQ2)}
Data sparsity issue is one of the main concerns of existing recommender systems.
More specifically, the recommendation performance would decrease
when the users' interactions become sparse.
Therefore, it is meaningful to study whether our propose SHGCN can still work well for those users with sparse interactions.
We study the performance of proposed model and baselines when it comes to different sparsity level. 
Specifically, we divide users  into several groups according to the number of purchase behaviour, and then evaluate models in different groups. We make sure each group has enough users or items to make the results stable. For each group, we report the averaged values of recommendation performance.

We present the Recall and NDCG metrics of both Beidian dataset and Beibei dataset in Figure \ref{fig::rq2::beidian} and Figure~\ref{fig::rq2::beibei}. 
To make it clear, we only show the two most competitive baselines, Diffnet and NGCF-BG.
As we can observe from the figures, our SHGCN achieve better performance in all four groups with different sparsity levels. This demonstrates the effectiveness of our model in alleviating the data-sparsity issue.

\vspace{0.2cm}
In short, our model show promising recommendation performance when faced with the data sparsity issue. 

\begin{figure*}[!t]
	\begin{center}
		\mbox{\subfloat{\includegraphics[scale=0.65]{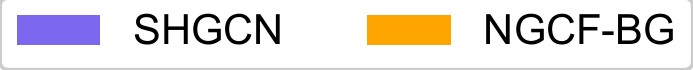}}}
	\end{center}
	\vspace{-0.8cm}
\end{figure*}
\begin{figure*}[!t]
	\begin{center}
		\mbox{
			% original: width=8.5
			\subfloat{\includegraphics[width=0.25\linewidth]{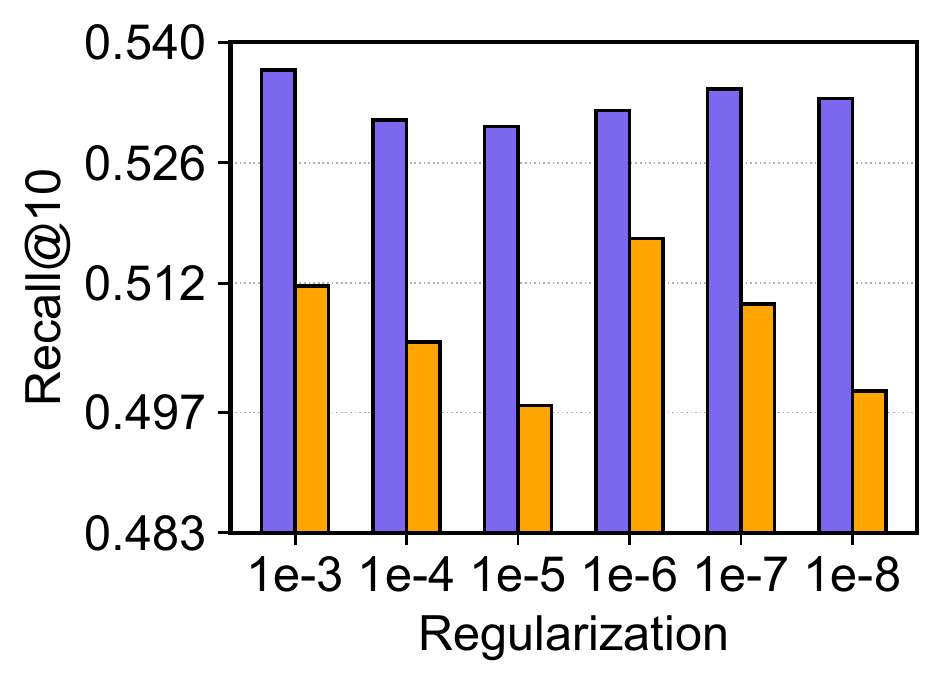}}
			\subfloat{\includegraphics[width=0.25\linewidth]{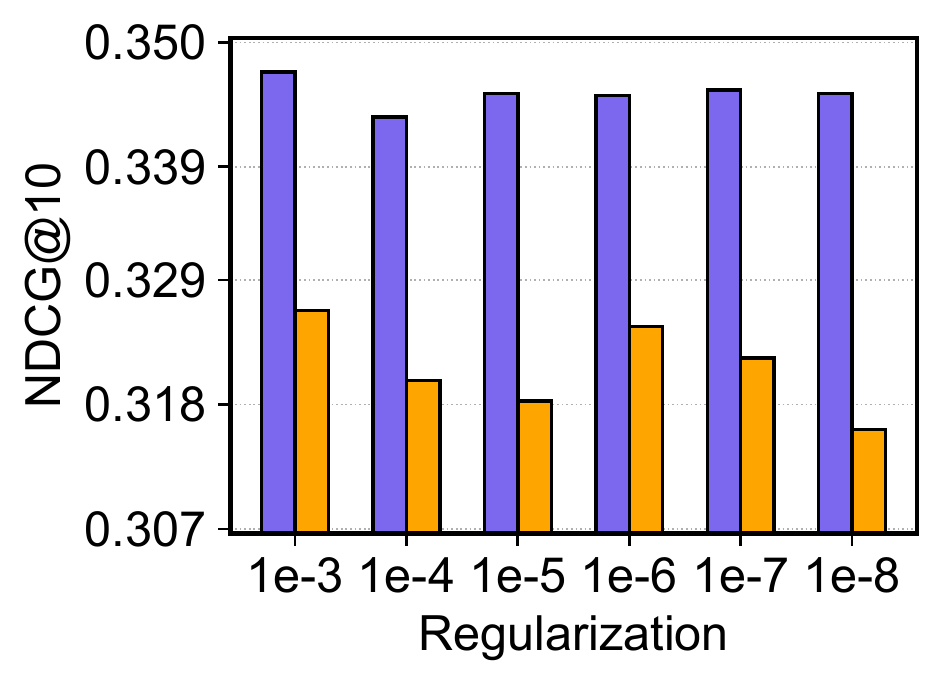}}
			\subfloat{\includegraphics[width=0.25\linewidth]{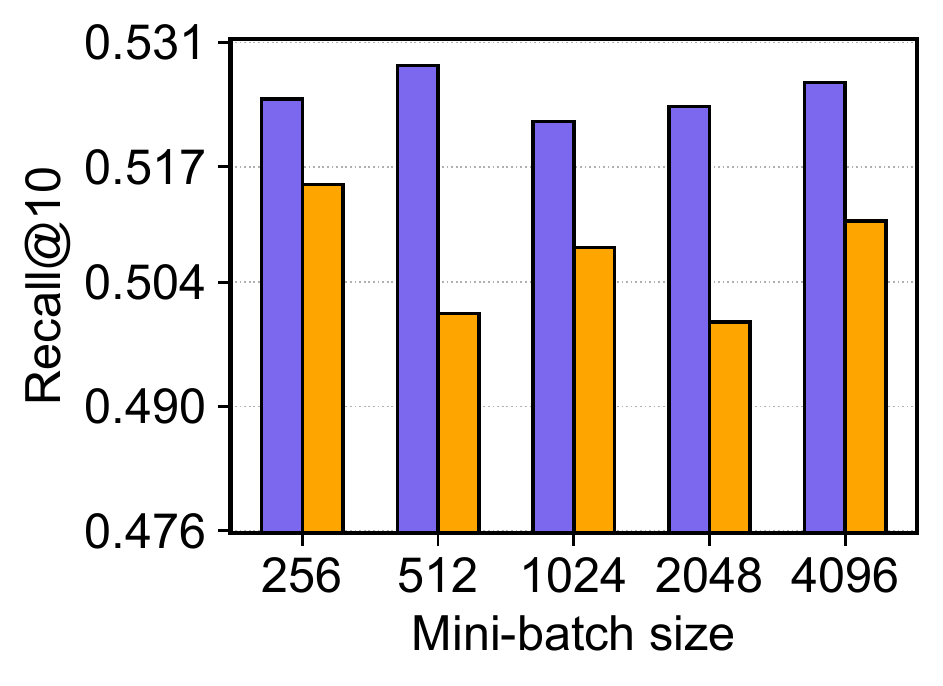}}
			\subfloat{\includegraphics[width=0.25\linewidth]{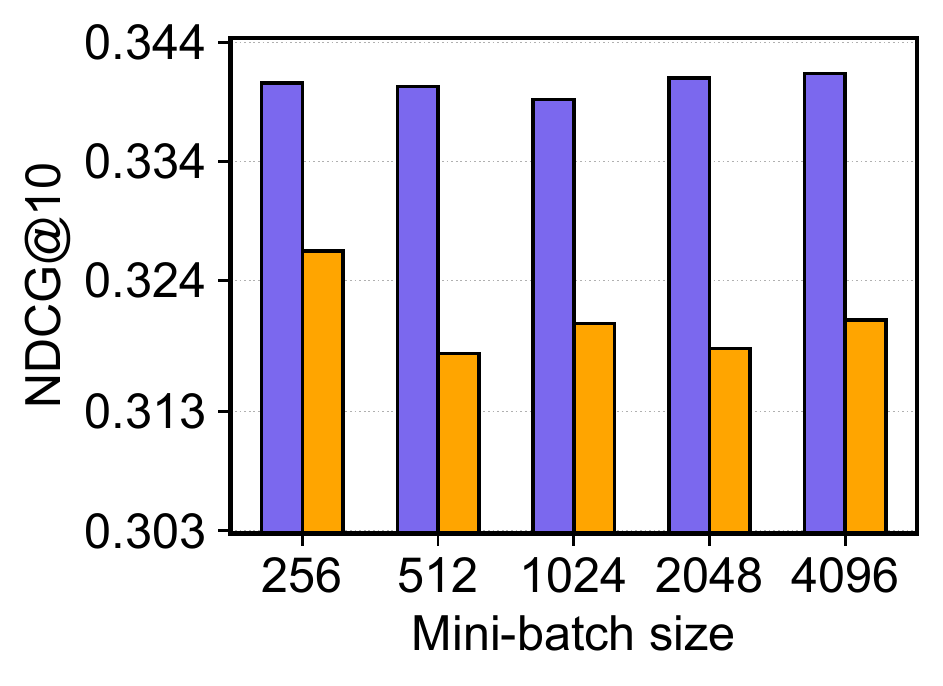}}
		}
	\end{center}
	\caption{Influence of hyper-parameters settings on the Beidian dataset.}\label{fig::rq3::beidian}
\end{figure*}

\begin{figure*}[!t]
	\begin{center}
		\mbox{
			% original: width=8.5
			\subfloat{\includegraphics[width=0.25\linewidth]{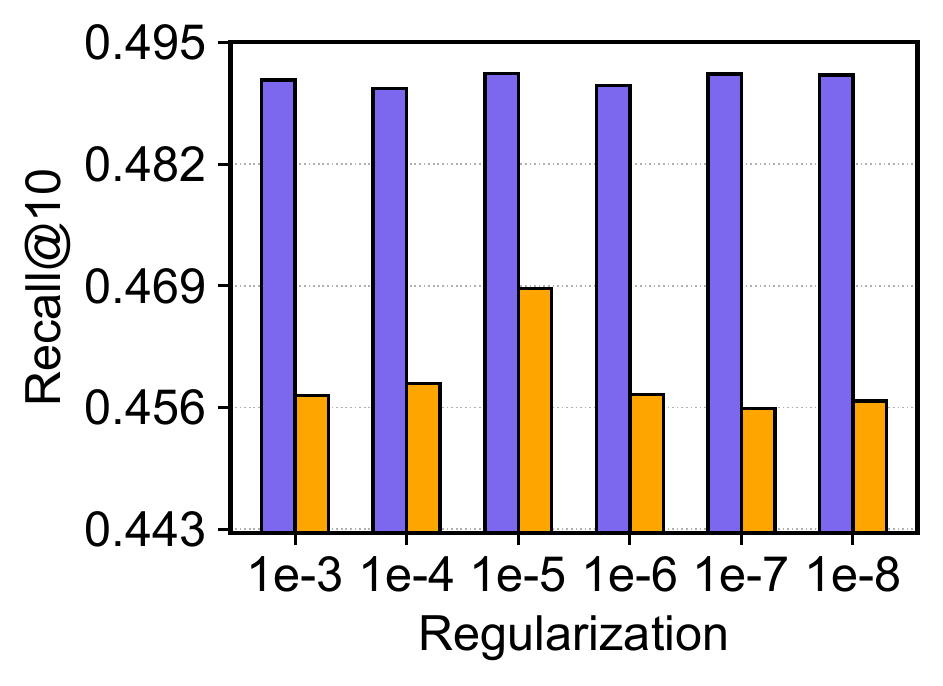}}
			\subfloat{\includegraphics[width=0.25\linewidth]{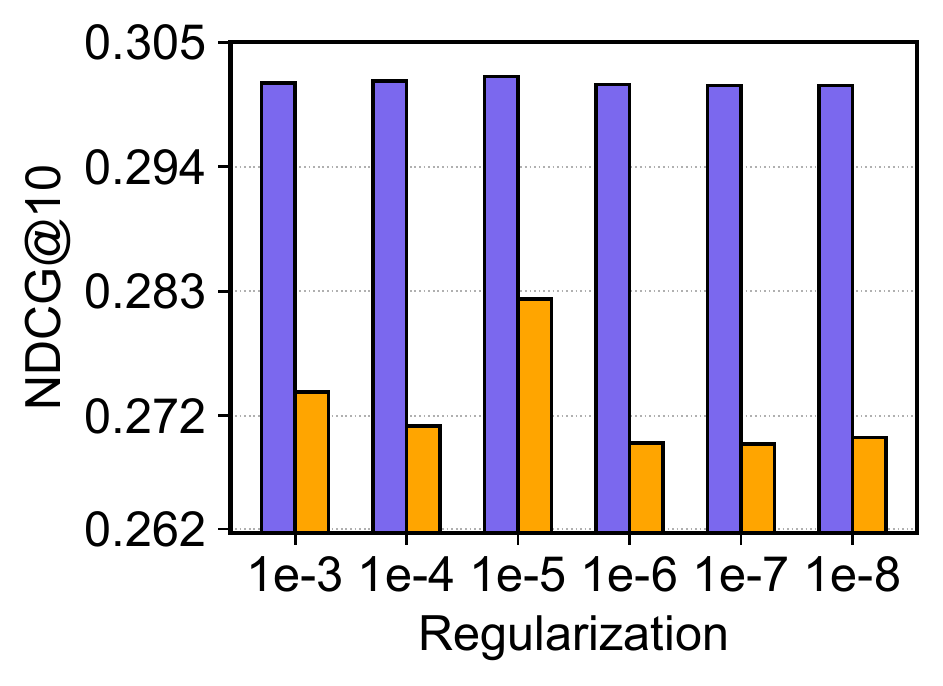}}
			\subfloat{\includegraphics[width=0.25\linewidth]{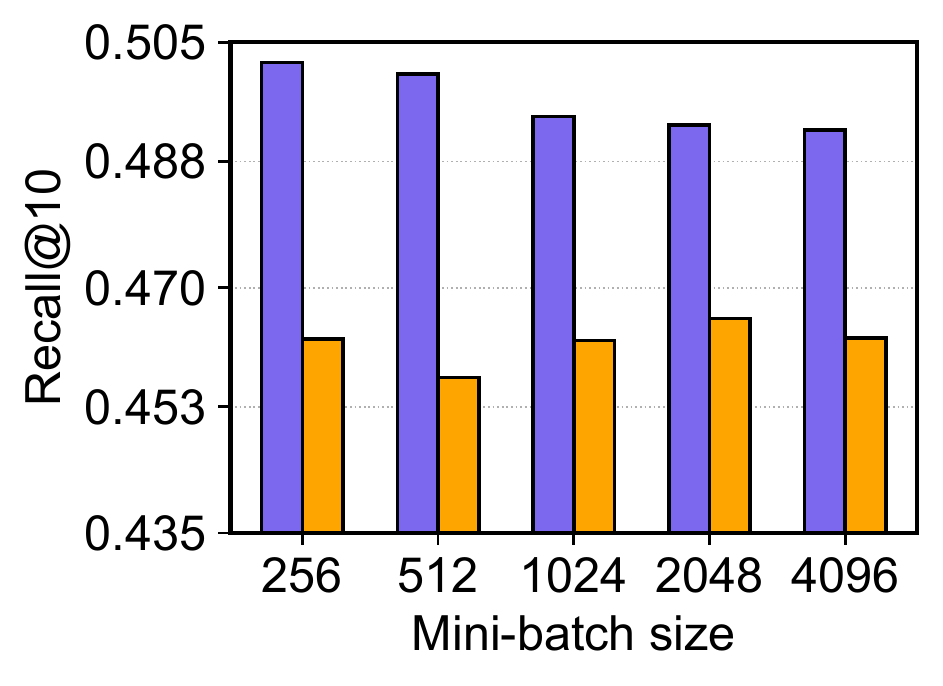}}
			\subfloat{\includegraphics[width=0.25\linewidth]{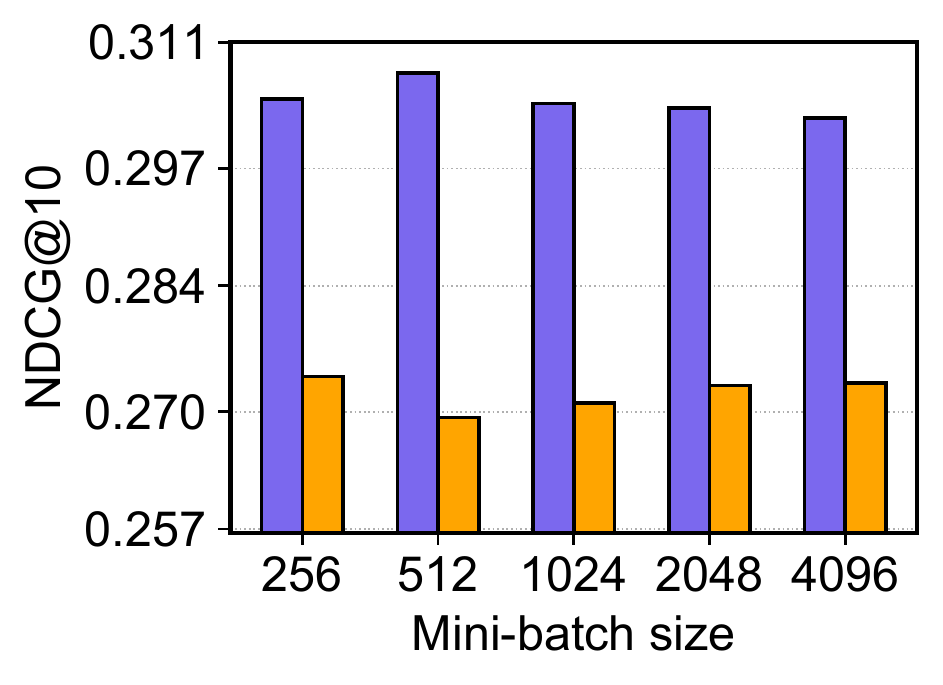}}
		}
	\end{center}
	\caption{Influence of hyper-parameters settings on the Beibei dataset.}\label{fig::rq3::beibei}
\end{figure*}

\subsection{Hyper-parameter Study (RQ3)}
Selecting proper hyper-parameters is always one of the most challenging issues for almost every deep learning models. 
Some models are very sensitive to the changes in hyper-parameters, which causes unstable model performance or costs a lot effots in tuning hyper-parameters.

In our experiment, we carefully search the learning rate and $L_2$ normalization coefficient as mentioned in Section \ref{subsectionsetting}. 
The empirical results show that our model outperforms all the  baselines under this setting. 
Furthermore, the setting of mini-batch size varies in \{256, 512, 1024, 2048, 4096\}.
These are three important hyper-parameters of our proposed model.
The detailed results with different choices of hyper-parameters are reported in Figure \ref{fig::rq3::beidian} and Figure \ref{fig::rq3::beibei}. From the results, we can observe that our SHGCN is not sensitive to the change of hyper-parameters. The fluctuation of NDCG@10 is limited to 1.16\% in the Beidian dataset and 2.07\% in the Beibei dataset when we vary the $L_2$ regularization term.
Besides, if we vary the mini-batch size, then the fluctuation of Recall@10 is limited to 1.20\% in the Beidian dataset and 1.94\% in the Beibei dataset.
These results show that our proposed model's performance is not sensitive to the setting of hyper-parameters. This promises its high practical application values in industrial scenarios.

\vspace{0.2cm}
In short, our proposed model is robust to the change of hyper-parameters and effort-saving in tuning hyper-parameters.
\section{Related Works}\label{sec::related}
We present the related works from two aspects: social recommendation and graph/hypergraph-based recommendation.

\para{Social Recommendation.} Social recommendation is generally defined as leveraging social-relational data to enhance recommendation systems. 
Since the user may have closer preferences with friends compared with strangers, social-relation data can help better capture users' preferences, which bringing better recommendation performance.
Existing works on social recommendation can be divided into two categories, social regularization and social smoothing.
Social regularization methods try to set constraints to friends' embedding distances. Some works design regularization terms in the objective function~\cite{soreg,jamali2010matrix,wang2017item} and some other works adopt multi-task learning on the preference learning and social-relation prediction~\cite{sorec,yang2017bridging}.
Actually, these works follow the common assumption of social trust, that friends tend to have similar preferences.

Some recent works start to study the more complex social relations in recommendation.
Yu~\textit{et al.}~\cite{yu2019generating} proposed that some friends do not share interests and design generative adversarial networks~\cite{goodfellow2014generative} to identifying these unreliable friends.
Chen~\textit{et al.}~\cite{chen2019social} model the different strengths of social relations with attention networks.
Although more complex social relations are considered, these works only use scalar weights to distinguish different strengths.
Therefore, the modeling of the social trust effect of existing works is still limited. Distinguishing different social relations merely by strengths is not enough to fully capture the social influence in the real world.
 
Compared with these works, in our work, we study social recommendation from a fine-grained perspective based on the user-item-user triple social relation.

\para{Graph-based and Hypergraph-based Recommendation.} Recommendation, in general, can be regarded as a kind of link-prediction task on the graph. Besides, the input data of recommendation system can be well organized by a graph structure, and thus graph-based models become the mainstream solution in today's recommender systems.

The early works~\cite{baluja2008video,gori2007itemrank,eksombatchai2018pixie,perozzi2014deepwalk} apply random-walk based methods and consider the visited nodes as the recommendation results. These works do not learn the latent representation of nodes, resulting in inferior recommendation results.
To improve them, some other works~\cite{chen2016query,yang2018hop,chen2019collaborative,tang2015line} adopt graph-embedding methods to learn embedding vectors of users and items in recommendation. The recommendation results are then obtained via embedding matching.
Yang~\textit{et al.}~\cite{yang2018hop} proposed to first capture user-item high-order connectivity with the graph structure and then introduce a matrix factorization model for recommendation.
Chen~\textit{et al.}~\cite{chen2019collaborative} utilized the neighboring nodes to calculate the similarity between users and items, and the similarity was further used for embedding learning in recommendation.
Recently, graph convolutional networks (GCN)~\cite{GCN, GAT, GIN} have become the state-of-the-art graph learning models, which can learn high-quality node embeddings.
The basis of GCN is conducting embedding propagation between neighbor nodes which can not only leverage the node features but also capture graph structure.
Due to their strong representation ability, GCNs are widely applied in recommendation tasks~\cite{ying2018graph, GCMC, wang2019neural, zheng2019price, wang2019kgat,wu2019session}.
Ying~\textit{et al.}~\cite{ying2018graph} applied GCN to pin-board graph in Pinterest with neighboring sampling for recommendation. 
Berg~\textit{et al.}~\cite{GCMC} proposed GCN-based model for rating-prediction recommendation.
Wang~\textit{et al.}~\cite{wang2019neural} proposed the GCN model for the general recommendation tasks. 
Besides collaborative filtering, GCNs also achieved great success in other recommendatoin tasks, such as session-based recommendation~\cite{wu2019session}, bundle recommendation~\cite{chang2020bundle}, knowledge-aware recommendation~\cite{wang2019kgat}, social recommendation~\cite{wu2019neural}, etc.

Recently, hypergraph representation learning has been developing rapidly and has been applied to recommender systems.
By generalizing the concept of edge to hyperedge, hypergraph can help resolve more complex relations compared with traditional graphs.
Van~textit{et al.}~\cite{van2017hypergraph} proposed a user-based collaborative filtering method for item recommendation incorporating information from social networks with hypergraph, which calculates the similarity of users with the hypergraph embedding. To smooth the latent factors in Low-rank matrix completion, Rao~\textit{et al.}~\cite{rao2015collaborative} filtered the latent factors with hypergraph regularized terms. Yu~\textit{et al.}~\cite{yu2019spectrum} proposed a spectral clustering-enhanced pairwise ranking method (SPLR) with the Laplacian matrix of the hypergraphs of users and items. Ji~\textit{et al.}~\cite{ji2020dual} proposed dual-channel hypergraph collaborative filtering(DHCF) to explicitly model the high-order correlations among users and items with a hypergraph.

\vspace{0.2cm}
In this work, we first construct the hypergraph representing triplet social relations as hyperedges and then propose a hypergraph convolutional network-based model. The model can learn the inhomogeneous social influence from the triple social relations via embedding propagation and aggregation.

\section{Conclusion and Future Work}\label{sec::conclusion}
In this paper, we approach the problem of inhomogeneous social recommendation, which studies the fine-grained social influence between friends.
We propose to construct a hypergraph that can well represent both the triple social relations and user-item interaction data.
We develop an effective hypergraph convolutional network model that effectively learns from the hypergraph.
Extensive experimental results on two real-world datasets demonstrate the effectiveness of our model. Further studies confirm that our model can help alleviate the data-sparsity issue significantly.
We also show that our model's recommendation performance is not sensitive to hyper-parameter settings.

For future work, we first plan to test our model's performance in more scenarios, such as group-buying in social networks, where more than two friends share one item.
Leveraging more auxiliary data, such as multi-relational information or explicit ratings of these complex interactions, is right on the drawing board to examine the effectiveness of our model on other tasks such as relation prediction or classification. We are also planning to take temporal information into consideration further to enhance the recommendation performance in more complicated applications.

\section{Acknowledgment}
This work is supported in part by National Natural Science Foundation of China (No. 62102420 and No.61832017), Beijing Outstanding Young Scientist Program NO. BJJWZYJH012019100020098, Intelligent Social Governance Interdisciplinary Platform, Major Innovation \& Planning Interdisciplinary  Platform for the ``Double\-First Class'' Initiative, Renmin University of China.

% \clearpage
\bibliographystyle{IEEEtran}
\bibliography{IEEE}
\balance
\end{document}